\documentclass[11pt]{article}

\usepackage[utf8]{inputenc}
\usepackage[T1]{fontenc}
\usepackage{amsmath,amssymb,amsthm}
\usepackage{mathtools}
\usepackage{graphicx}

\usepackage{xcolor}

\graphicspath{{./}}
\usepackage[margin=1in]{geometry}
\usepackage{hyperref}
\usepackage{caption}

\newcommand{\Op}{\mathrm{Op}}
\newcommand{\Ass}{\mathcal{A}}

\newcommand{\Rtwo}{\mathbb{R}^{2}}
\newcommand{\phibar}{\bar{\phi}}

\theoremstyle{definition}

\newtheorem{condition}{Condition}
\newtheorem{remark}{Remark}

\title{A heterogenetic theory of living systems}
\author{Alessandro Sarti\thanks{CAMS, CNRS-EHESS, Paris. \texttt{alessandro.sarti@ehess.fr}} \and
        Giovanna Citti\thanks{Dipartimento di Matematica, Universita di Bologna. \texttt{giovanna.citti@unibo.it}}}

\date{}

\begin{document}

\maketitle

\begin{abstract}
The problem of the dynamics of living systems and their mathematical representation essentially unfolds on two levels. Firstly, the principles of autopoiesis and operational closure describe how a living system produces and maintains itself. Secondly, living phenomena are recognised as historical, involving the evolution of dynamics and spaces of possibility that change over time ~\cite{longo2014,longo2017}. Currently, no differential model can incorporate both of these aspects. We propose adopting the differential heterogenesis framework~\cite{sarti2018,sarti2022}, which enables us to define differential operators that change in space and time, producing event-based and historical dynamics. In this construction, heterogeneous operator assemblages will be constrained by the principles of autopoiesis and operational closure, providing a theoretical framework for event-based and historical living systems. Finally, we provide a specific example of assembly with Gray-Scott kinetic equations, along with numerical simulations and results.
\end{abstract}

\section{Introduction}

The problem of the dynamics of living systems and their mathematical representation unfolds at least on two levels. Firstly, there are the principles of autopoiesis and operational closure, which describe how a living system produces and maintains itself. On the other, living phenomena are historical and the possibility spaces in which they unfold change over time. 

\subsection{Autopoiesis and operational closure}

In the twentieth century, the question of what distinguishes living systems from physical and chemical ones was reformulated as a search for the minimal organisation that qualifies a system as living. The concept of autopoiesis, introduced by Maturana and Varela ~\cite{maturana1980,varela1974}, describes a living system as a network of production processes that continuously regenerate the components which sustain the network. This organisation is both the producer and the product of its own activity, and the boundary that separates it from the environment is produced from within. In this view, a cell is a process that synthesises its own membrane, enzymes, and genetic machinery, using these very structures as the means of synthesis.

Related programmes have expressed the same organisational logic in formal
terms. Rosen's relational biology~\cite{rosen1991} characterises
self-maintenance as closure to efficient causation, every functional component
being produced by some other component within the system, which amounts to a
fixed-point property at the level of functions. Kauffman's theory of
autocatalytic sets~\cite{kauffman1986,kauffman1993,kauffman2000,hordijk2004}
approaches the same question from the origin of life, showing that in
sufficiently rich chemical spaces self-sustaining reaction networks arise
generically. Mont\'evil and Mossio~\cite{montevil2015,mossio2016} have
developed this tradition into a theory of organisational closure, in which a
system is closed when it exhibits a network of constraints, such as enzymes,
membranes or regulatory circuits, each of which is regenerated by processes
depending on the others. In all these accounts closure takes the form of a
fixed-point condition.

\subsection{Historicity and the change of spaces of possibility}

Longo and Mont\'evil have argued that the mathematical structures of physics,
with their fixed state spaces and invariant laws, are inadequate to biological
evolution and development~\cite{longo2014,longo2017}. Living systems do not
simply traverse a pre-given space of possibilities, since they construct new
dimensions of variation as they evolve.  Therefore, a biological object is defined by its history: the partly contingent sequence of events that produced it and that it carries forward. Such a history is transmitted through reproduction with variation; each generation reproduces the organisation that it has developed, ensuring that any novelty introduced at one stage is conserved and made available at the next.
The two levels pull in different directions. Closure is a fixed-point condition
and seems to require invariance, while historicity requires that the space in
which the fixed point is sought be allowed to change.

\subsection{Heterogenesis as a theory of operator assemblages}

To hold the two levels together we adopt the framework of differential
heterogenesis developed by Sarti, Citti and Piotrowski~\cite{sarti2018,sarti2022}.
Its primitive object is not a single differential  operator acting on a function defined in a certain domain but rather an assemblage of heterogeneous differential operators changing in space and time. 

Heterogenesis introduces then a double temporality. Along  with the the time $t$, which is the time along
which solutions evolve under the operators, there is the time $\tau$, which is the time of
events, along which the operators themselves are assembled, varied and
transmitted. In physics only the first is present, since forms may change but
the laws that generate them are invariant. The temporality $\tau$ allows the
operators to change, and each such change is an event in the proper sense. The
space of possibilities is modified along $\tau$, and historicity can then be
understood as an organisation of events.

\subsection{A heterogenetic theory of living systems}

The present paper aims to apply the formulation of differential heterogenesis to theoretical biology by incorporating the constraints associated with the autonomy of living systems into operator assemblages.

In this framework a level-set function $\phi$ individuates a body, whose interface
$\Gamma=\{\phi=\phibar\}$ is a free boundary produced by the dynamics it
delimits. Two heterogeneous operators act on the two sides of $\Gamma$, forming
a spatial assemblage, and these operators change along $\tau$, forming a
temporal assemblage. On this object we impose a closure of constraints, which is a fixed
point in the space of fields, and a
reproduction of the assemblage, expressed as a closed orbit in phase space.
We will show that reproduction with operators changing along $\tau$ correspond to reproduction with variation. Historicity follows from the construction.
In this setting a cell becomes a laboratory of operator assemblage, and each unit is
defined by the sequence of events its ancestors organised and transmitted.

What we propose is a theory and not a model. Its purpose is to indicate how
pertinent models of living systems can be constructed, by specifying the
objects involved and the constraints they must satisfy. For this reason the
equations of a specific model appear only at the end of the paper, as an
example. 

The paper is organized as in the following: Section~\ref{sec:problem} sets up the heterogenetic problem, with the definition of the 
level-set function, the interface $\Gamma$ and the spatial assemblage.
Section~\ref{sec:conditions} defines the formal spaces involved and states the closure
of constraints and the reproduction condition. Section~\ref{sec:historicity}
develops the temporal assemblage and the historicity that follows from it.
Section~\ref{sec:gs} and Section~\ref{sec:numerics} present respectively the assemblage equations with specific Gray-Scott operators and related numerical simulations , while
Section~\ref{sec:discussion} discusses the main point of the theory. 

\section{The heterogenetic framework}
\label{sec:problem}

Let $\Omega \subset \Rtwo$ be a bounded reference domain and let
\[
\phi : [0,T] \times \Omega \longrightarrow \mathbb{R}, \qquad (t,x) \mapsto \phi(t,x),
\]
be a level-set function. Fixed a threshold value
$\phibar \in \mathbb{R}$, at each time $t$ the sign of $\phi-\phibar$ induces a
partition of the domain into two phases.

The interior phase is the super-level set
\[
\Omega_i(t) = \bigl\{\, x \in \Omega : \phi(t,x) \ge \phibar \,\bigr\},
\]
a simply connected subdomain of $\Rtwo$, and the exterior phase is its
complement, the sub-level set
\[
\Omega_o(t) = \bigl\{\, x \in \Omega : \phi(t,x) < \phibar \,\bigr\}.
\]
The free boundary $\Gamma(t)$ is the level set of $\phi$ at the threshold
$\phibar$,
\[
\Gamma(t) \;=\; \bigl\{\, x \in \Omega \;:\; \phi(t,x) = \phibar \,\bigr\}.
\]

On the two phases we let act two heterogeneous differential operators,
$\Op_i$ on the interior $\{\phi\ge\phibar\}$ and $\Op_o$ on the exterior
$\{\phi<\phibar\}$. The heterogeneity of the two operators is what
makes the problem \emph{multiphase}.

The complete system reads
\begin{equation}
\label{eq:system}
\left\{
\begin{aligned}
\Op_i^{\tau}\bigl(\phi(t,x)\bigr) &= 0, && \phi(t,x) \ge \phibar, \\[2pt]
\Op_o^{\tau}\bigl(\phi(t,x)\bigr) &= 0, && \phi(t,x) < \phibar, \\[2pt]
\Gamma(t) &= \bigl\{\, x : \phi(t,x) = \phibar \,\bigr\}, && \\[2pt]
\phi(0,x) &= \phi_0(x), && x \in \Omega,
\end{aligned}
\right.
\end{equation}
where $\phi_0$ is a prescribed initial field. The first two equations encode
the two phases, defined respectively on the super-level set $\{\phi\ge\phibar\}$
and the sub-level set $\{\phi<\phibar\}$; the third states explicitly that the
interface is the $\phibar$-level set of the solution---the free-boundary
coupling---and the last fixes the initial datum.
Since $\Gamma(t)$ is determined by the solution $\phi$ itself, the interface is not known in
advance, and the problem is therefore a free-boundary problem.

Two distinct temporalities are at work here, and the notation separates them.
The variable $t$ is the temporality of the solution $\phi$, the time along which the field evolves, the interface moves and the phases change shape.
The variable $\tau$ is the temporality of the operators, the time of the events
along which the operators themselves are assembled, varied and transmitted.

Writing $\Op_i^{\tau}$ and $\Op_o^{\tau}$ expresses this distinction, since the operators acting on
the two phases are not fixed, and the system \eqref{eq:system} is to be read at fixed $\tau$ as a
free-boundary problem in $t$, the operator pair being what slowly changes with
$\tau$. When the operators are constant in $\tau$ the two times collapse and
\eqref{eq:system} reduces to an ordinary free-boundary multiphase problem. 
The case of interest is the one
in which they vary.\footnote{ If the discontinuity between $\Op_i$   and $\Op_o$ across Gamma is problematic, one can work directly with the assemblage
\[
\Op \;=\; \sigma\,\Op_i^{\tau} \;+\; (1-\sigma)\,\Op_o^{\tau},
\]
where $\sigma=\sigma(\phi-\phibar)$ is a sigmoid centred on $\Gamma$. The pair
$\{\sigma,1-\sigma\}$ is then a partition of unity subordinate to the two
phases, and the passage from one operator to the other becomes a smooth
transition across the interface rather than a jump; the sharp system
\eqref{eq:system} is recovered in the limit of an infinitely steep sigmoid.}

We call \emph{assembly}, and denote by $\Ass(\tau)$, the system
\eqref{eq:system} itself, that is the two operators together with the boundary condition
acting through the free boundary. The assembly is indexed by $\tau$ because it
is the pair of operators as it stands at that moment of its history that
constitutes it. The dependence on $\tau$ of the assemblage is what carries the historical dimension.

Acting on a field $\phi$, the assembly returns the field produced by the system
\eqref{eq:system}; we write this action as $\Ass\bigl(\phi(x)\bigr)$. The
assembly simultaneously determines the interface, through
$\Gamma = \{x : \phi(x) = \phibar\}$, so that $\Ass$ acts on $\Gamma$ as well
and we write $\Ass(\Gamma)$ for the induced map on the boundary. 

\begin{figure}[htbp]
\centering
\includegraphics[width=0.72\textwidth]{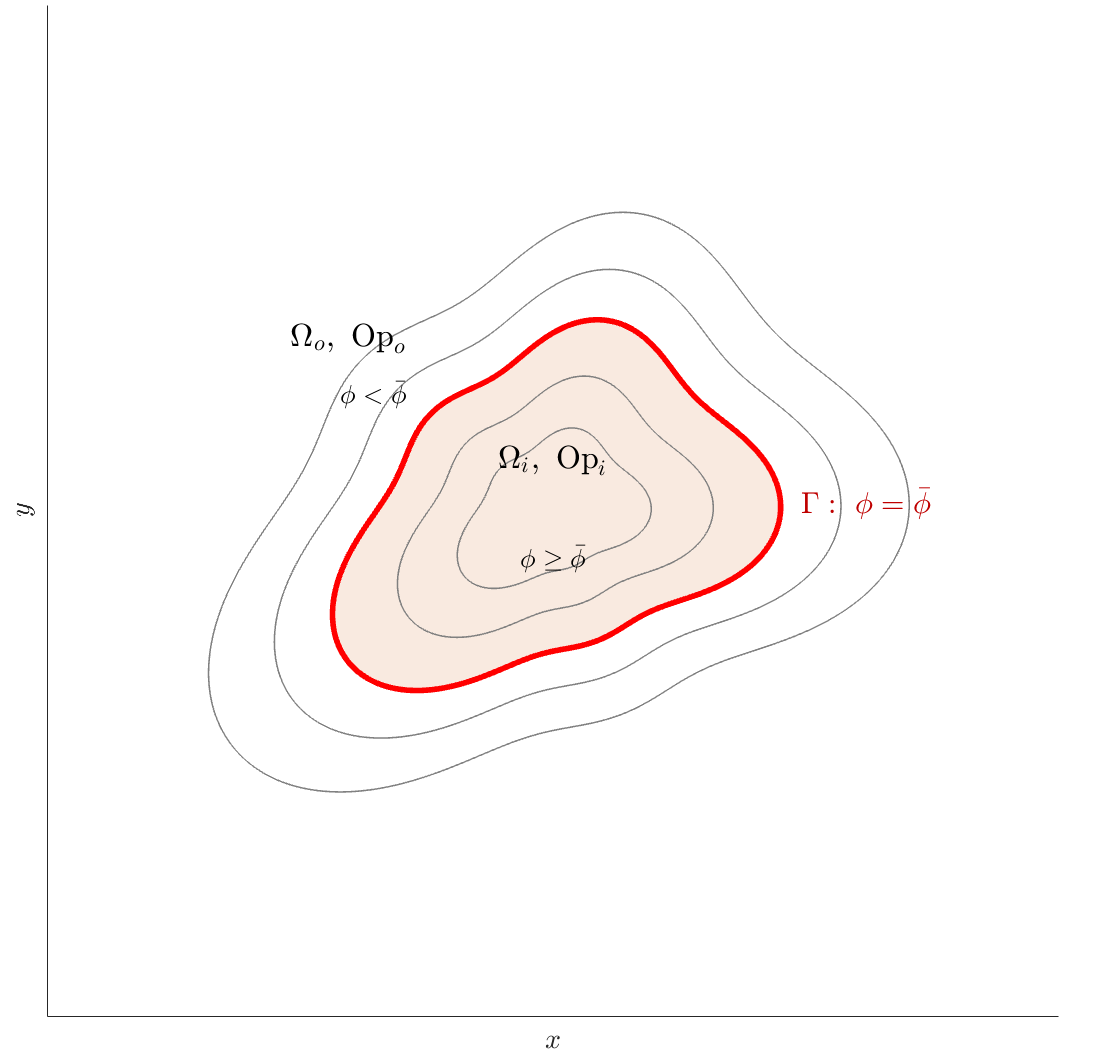}
\caption{The level-set function $\phi$ realised as a two-dimensional
spot, seen from above. The grey curves are level sets of $\phi$; the thick red
curve is the free boundary $\Gamma=\{x:\phi(x)=\bar\phi\}$, which is
recovered as the level set at the threshold $\bar\phi$. Inside it lies the
interior phase $\Omega_i=\{\phi\ge\bar\phi\}$, governed by
$\mathrm{Op}_i$; outside it the exterior phase
$\Omega_o=\{\phi<\bar\phi\}$, governed by $\mathrm{Op}_o$.}
\label{fig:assembly}
\end{figure}

Figure~\ref{fig:assembly} shows the reference configuration, with the level curves of
$\phi$, the free boundary $\Gamma$, and the two phases it separates. The labels
$\Op_i$, $\Omega_o$, $\Op_o$ and $\Gamma$ identify the two operators, the outer
phase, and the interface.

Without the introduction of further constraints, the integration of this differential assembly can give rise to extremely varied spatiotemporal dynamics which, in extreme cases, can be completely immanent and anarchic. We therefore introduce the specific conditions inherent in the dynamics of living organisms in order to limit their excess degrees of freedom.

\section{Closure and reproduction}
\label{sec:conditions}

The two conditions that organise the construction are best introduced by
separating the objects they act upon, because they do not live in the same
formal space. Two levels are involved.

\subsection{Functional spaces}

Let $X$ be the Banach space of fields,
\begin{equation}
\label{eq:space-X}
X \;=\; C^{0}\bigl([0,T];L^{2}(\Omega)\bigr)\;\cap\;L^{2}\bigl([0,T];H^{1}_{0}(\Omega)\bigr),
\end{equation}
and let $\mathcal{O}$ be the space of operator pairs,
\begin{equation}
\label{eq:space-O}
\mathcal{O} \;=\; \Bigl\{\,(\Op_i,\Op_o)\;:\;
g_i,g_o\in W^{1,\infty}(\Omega),\quad g_i,g_o\ge\gamma_0>0,\quad
\|g_i\|_{W^{1,\infty}},\|g_o\|_{W^{1,\infty}}\le M \,\Bigr\},
\end{equation}
a closed bounded subset of $W^{1,\infty}(\Omega)^{2}$.    The ellipticity bound
$g\ge\gamma_0>0$ keeps the operator uniformly
non-degenerate and allows the system \eqref{eq:system} to be solved.  Weaker conditions can ensure existence of solutions as for example sub-ellepticity~\cite{R-S1976}.   The entry
$(\Op_i^{\tau},\Op_o^{\tau})$ of the assembly is an element of $\mathcal{O}$. 

Two maps act on these objects. The first is
\begin{equation}
\label{eq:map-A}
\Ass^{\tau} : X \longrightarrow X,
\end{equation}
the assembly at fixed $\tau$, which takes a field, solves the free-boundary
problem, and returns the field. The second is the cycle (closed orbit)  in
phase space,
\begin{equation}
\label{eq:map-C}
C_{\Ass}^{\tau} : \mathcal{P} \longrightarrow \mathcal{P},
\qquad
\mathcal{P}=\bigl\{(\phi,\nabla\phi)\;:\;\phi\in X\bigr\},
\end{equation}
generated by the assembly $\Ass^{\tau}\in\mathcal{O}$. Its argument is a point
of phase space and the assembly is its generator, which explains the subscript,
while the superscript $\tau$ records the time of the operators.

\subsection{Closure of constraints}

Since the pair carries its own time $\tau$, the assembly is
$\Ass^{\tau}=(\Op_i^{\tau},\Op_o^{\tau})$, and two readings must be
distinguished.

\emph{Fixed $\tau$.} Freeze the pair at $\tau=\bar\tau$. Then \eqref{eq:system}
is an ordinary free-boundary problem whose only unknown is the field. The
interface $\Gamma$ is then a constraint that the field must satisfy and which the
field in turn regenerates.

\begin{condition}[Closure of constraints]
\label{cond:fixed}
For certain values $\bar\tau$ of the operator time a field $\phi$ is a fixed
point of the assembly frozen at $\bar\tau$, if
\begin{equation}
\label{eq:fixed}
\Ass^{\bar\tau}\bigl(\phi\bigr) = \phi .
\end{equation}
Consequently the free boundary is invariant,
\begin{equation}
\label{eq:fixed-gamma}
\Gamma\bigl(\Ass^{\bar\tau}(\phi)\bigr) = \Gamma(\phi).
\end{equation}
Equation \eqref{eq:fixed-gamma} is not independent, since it follows from
\eqref{eq:fixed} through the trace map $\Gamma(\cdot)$. This is the differential
form of the \emph{closure of constraints} of Mont\'evil and
Mossio~\cite{montevil2015,mossio2016}, in which the interface is stable on the timescale
of the processes it canalises, and is regenerated by them.
\end{condition}

Existence of a solution of \eqref{eq:fixed} follows from a Schauder-type
argument, since $\Ass^{\bar\tau}$ is nonlinear and non-local but maps a closed
bounded convex set into a compact subset of itself. Uniqueness is not expected,
and is not needed.

\emph{Slowly varying $\tau$.} Left free to vary, the same condition no longer
selects isolated fields but traces a curve in the space of fields,
\begin{equation}
\label{eq:fixed-family}
\Ass^{\tau}\bigl(\phi(\cdot,\tau)\bigr) = \phi(\cdot,\tau),
\qquad \tau\in[0,\tau_{*}],
\end{equation}
the interface following as $\Gamma(\tau)=\{x:\phi(x,\tau)=\phibar\}$. The
constraint is closed at every instant, and what the system traverses is a
one-parameter family of closures. A trace of history is thus already present at
this level, before reproduction (Condition~\ref{cond:repro}) is imposed at all.

\subsection{Reproduction}

\begin{figure}[htbp]
\centering
\includegraphics[width=\textwidth]{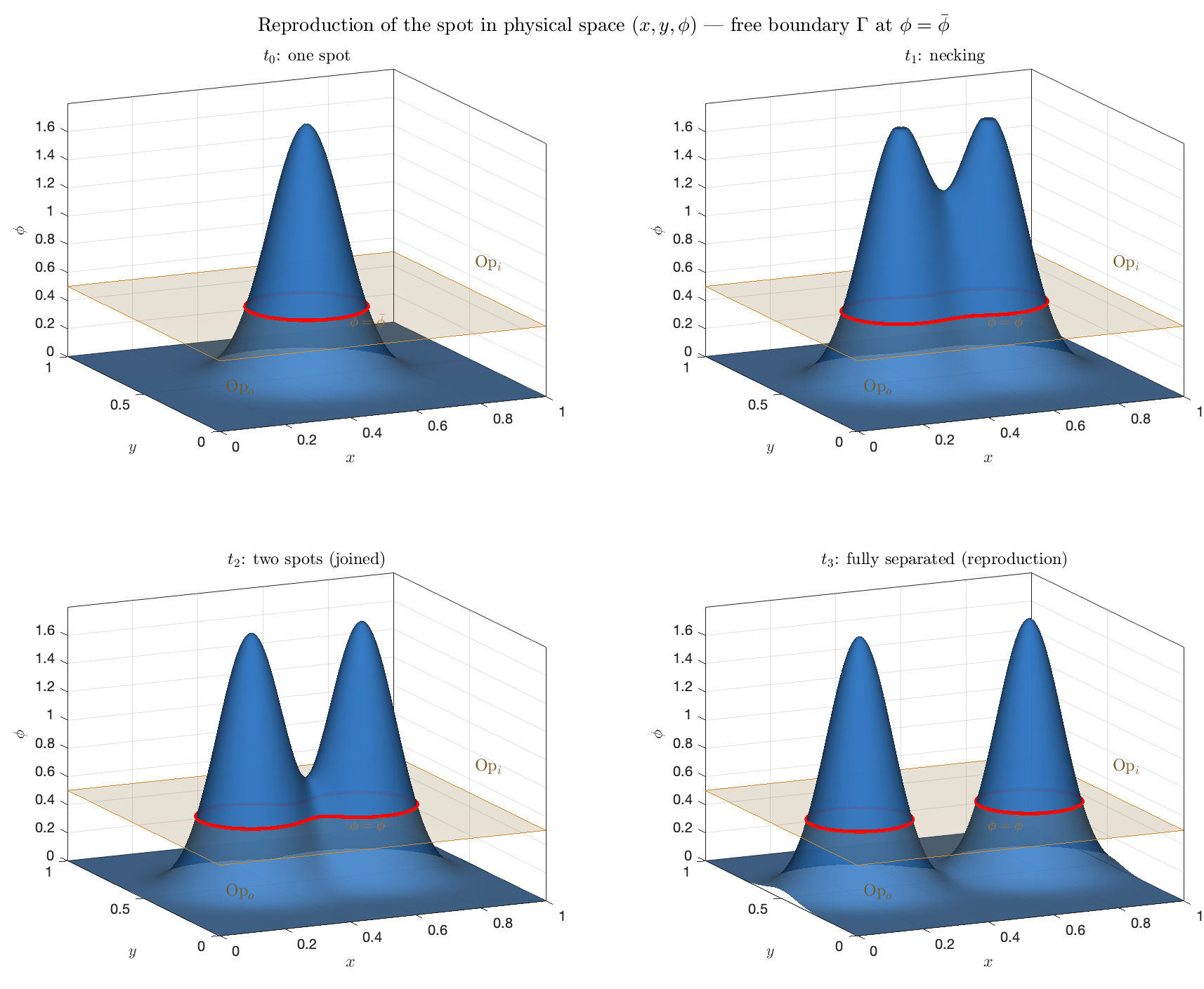}
\caption{Reproduction of the assembly in physical space $(x,y,\phi)$ at four
instants. At $t_0$ there is a single spot, at $t_1$ it necks, at $t_2$ two spots remain
joined at the threshold and at $t_3$ they are fully separated. The surface is the
level-set function $\phi(x,y)$, the semitransparent plane is $\phi=\phibar$,
and the red curve is the free boundary
$\Gamma=\{x:\phi(x)=\bar\phi\}$. Inside $\Gamma$ lies the interior phase
$\Omega_i=\{\phi\ge\phibar\}$ governed by $\Op_i$; outside it the exterior phase
$\Omega_o=\{\phi<\phibar\}$ governed by $\Op_o$. Reproduction is read at once as
the doubling of the spot and as the doubling of the closed curve $\Gamma$.}
\label{fig:spots-xy}
\end{figure}

\begin{figure}[htbp]
\centering
\includegraphics[width=\textwidth]{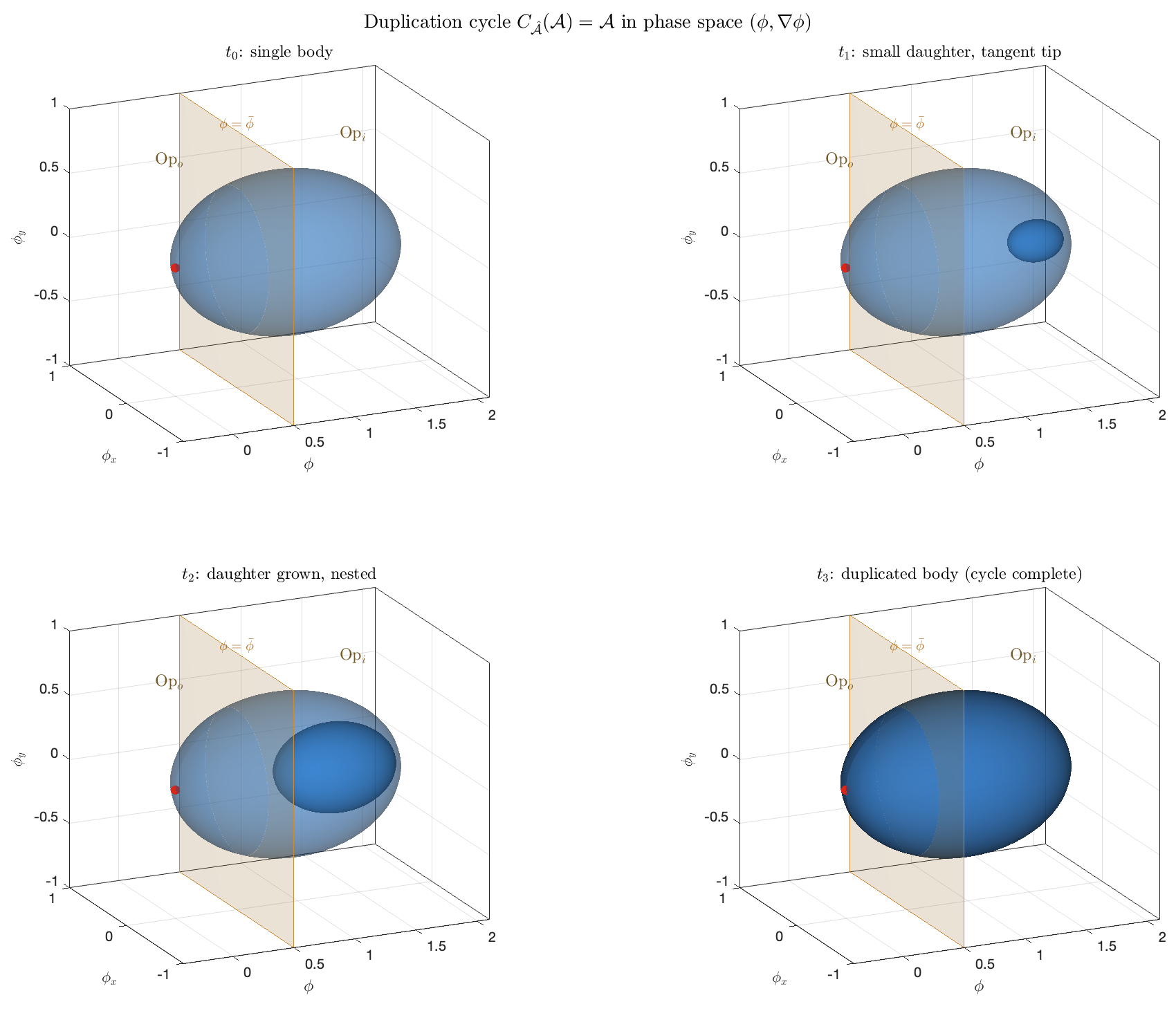}
\caption{The duplication cycle rendered in the phase space
$(\phi,\nabla\phi)$ as bodies of revolution. Each leaf-shaped orbit of the
level-set field is revolved about the $\phi$-axis into an ellipsoid-like body;
the semitransparent plane $\phi=\phibar$ is the free boundary $\Gamma$, with
the exterior operator $\Op_o$ acting where $\phi<\phibar$ and the interior
operator $\Op_i$ where $\phi\ge\phibar$; the red dot marks the fixed point at
the origin. From $t_0$ (a single assembly) a daughter nucleates tangentially at
the high-$\phi$ tip ($t_1$), grows in place ($t_2$), and matures into a
congruent copy, so that $t_3$ shows the duplicated body closing the cycle.}
\label{fig:cycle}
\end{figure}

The second condition is the one of reproduction, which allows the system to produce from a
flow $\phi$ a new flow, similar to the previous one but possibly displaced in space (Figure~\ref{fig:spots-xy}). To identify the invariances
of this process we represent the flow in the phase space of the variables
$(\phi,\nabla\phi)$. More precisely, we consider the section of the graph
$(x,\phi,\nabla\phi)$ in the space of the variables $(\phi,\nabla\phi)$. The
position $x$ is forgotten, and two spots that are copies of one another,
wherever they lie in the domain, are sent to the same closed curve
(Figure~\ref{fig:cycle}). The cycle $C_{\Ass}^{\tau}$ is the map of phase space
into itself generated by the assembly $\Ass^{\tau}$. Its argument is a point
$(\phi,\nabla\phi)$, and the subscript records that it is the assembly that
generates the cycle.

\begin{condition}[Reproduction of the assembly]
\label{cond:repro}
In certain other time regimes the assembly is reproduced by a closed orbit in
phase space,
\begin{equation}
\label{eq:repro}
C_{\Ass}^{\tau}\bigl(\phi,\nabla\phi\bigr) = \bigl(\phi,\nabla\phi\bigr).
\end{equation}
\end{condition}
We can now extend the operator $C_{\Ass}^{\tau}$ to an operator $\hat{C}_{\Ass}^{\tau}$
acting on the space of assemblages.

Indeed we can univocally associate to a point  $(\phi,\nabla\phi)$ of phase space 
the assembly 
$\Ass^{\tau}=(\Op_i^{\tau},\Op_o^{\tau},\{\phi=\phibar \text{ on } \Gamma\})$.
In facts, the interface $\Gamma$ is not an independent datum, being the level set
$\{x:\phi(x)=\phibar\}$, hence a function of $\phi$ alone,
$\Gamma=\Gamma(\phi)$. Consequently the boundary condition $\phi$ in the system \eqref{eq:system}, and the state $(\phi,\nabla\phi)$ determine
the assembly that acts on it. There is therefore a well-defined map
$\pi:\mathcal{P}\to\mathcal{O}$, defined on the phase space $\mathcal{P}$ with values in the 
space of assemblages, which associates
$(\phi,\nabla\phi)$ to the corresponding assemblage $\Ass^{\tau}(\phi)$.
Viceversa, we can associate to an assemblage 
$\Ass^{\tau}$ the couple $(\phi,\nabla\phi)=\pi^{-1}(\Ass^{\tau})$, 
Composing with $\pi$, we can define the operator 
$$\hat{C}_{\Ass}^{\tau}\bigl(\Ass^{\tau}\bigr)=  \pi\bigl(C_{\Ass}^{\tau}(\pi^{-1}(\Ass^{\tau}))\bigr)$$
Consequently 
equation \eqref{eq:repro} is equivalent to the reproduction of the assembly,
\begin{equation}
\label{eq:repro-assembly}
\hat{C}_{\Ass}^{\tau}\bigl(\Ass^{\tau}\bigr) = \Ass^{\tau}.
\end{equation}

Equations \eqref{eq:repro} and \eqref{eq:repro-assembly} thus express the same
condition from the two sides of the map $\pi$. The first states it in phase
space, where the orbit is visible as the closed curve of
Figure~\ref{fig:cycle}; the second states it in the space of operators, where
its content becomes explicit, since the two operators $\Op_i^{\tau}$ and $\Op_o^{\tau}$,
together with the interface they generate, reappear in the duplicate with the
same operators and not only the same outline of $\Gamma$. The orbit does not
only return the field to itself, it instantiates the assembly twice,
as shown in Figure~\ref{fig:spots-xy} in physical space and in
Figure~\ref{fig:cycle} in phase space.

As at the first level, two readings must be distinguished. At fixed $\bar\tau$
the orbit is closed and the assembly is reproduced identically. If the operators
vary slowly in $\tau$, the orbit no longer closes on the same point but on a
nearby one, and the duplicate carries $\Ass^{\tau'}$ with $\tau'\neq\tau$. The
cycle opens into a spiral, and reproduction becomes reproduction with
variation. This is developed in Section~\ref{sec:historicity}.

\begin{remark}[On the cycle $C_{\Ass}^{\tau}$]
The phase-space cycle is left deliberately generic in the present treatment. A
concrete realisation is furnished by the specific choice of operators. For example reaction--diffusion dynamics of
Gray--Scott type will be introduced in Section 5, whose closed orbits in $(\phi,\nabla\phi)$ provide an explicit
$C_{\Ass}^{\tau}$. But here we do not commit to that particular dynamics
and regard $C_{\Ass}^{\tau}$ as representing an unspecified closed orbit. The two
levels then describe complementary regimes, the closure of constraints
(Condition~\ref{cond:fixed}) and the reproduction of the assembly
(Condition~\ref{cond:repro}).
\end{remark}

\section{Time assemblages and historicity}
\label{sec:historicity}

The two conditions have a consequence that goes beyond the mechanics of the
equations and concerns the biological character of the construction. Each variation of the assemblage is an event in the time $\tau$. It may occur at duplication, when the daughter is built with an operator its
mother did not have, or in the course of the life of a single cell, which modifies its own operators while persisting as a unit. 

\subsection{Variation as temporal assemblage}

Such a variation does not simply replace one operator by another, but places the
operators in a temporal assemblage.
If $\Op_{i1}$ is the old interior operator and $\Op_{i2}$ the newly invented
one, the cell does not pass from the first to the second but comes to be
governed by their assemblage
\begin{equation}
\label{eq:time-assemblage}
\Op_i \;=\; a_1\,\Op_{i1} \;+\; a_2\,\Op_{i2},
\end{equation}
in which the coefficients $a_1,a_2$ do not blend the two operators into an
average but select which of them is canalised. They act as selectors, taking
values in $\{0,1\}$ or close to them, so that at each
instant the dynamics is channelled through one operator or the other according
to the occasion, while the operator momentarily left aside is not destroyed but
merely de-selected. Substitution would mean discarding $\Op_{i1}$ altogether;
here $\Op_{i1}$ remains in the assemblage, available to be selected again.

More generally, after a succession of $n$ inventions the interior operator is
the temporal assemblage
\begin{equation}
\label{eq:time-assemblage-n}
\Op_i(t) \;=\; \sum_{k=1}^{n} a_k(t)\,\Op_{ik},
\qquad a_k(t)\ge 0,
\end{equation}
and likewise for $\Op_o$. In \eqref{eq:time-assemblage-n} all $n$ operators remain permanently
available. None is destroyed by the arrival of its successor, and
each can be re-selected whenever the occasion requires it, simply by a change
in the corresponding coefficient $a_k$.    

The spatial assemblage of Section~\ref{sec:problem}, made of two
heterogeneous operators glued across the interface $\Gamma$, is thus
completed by a temporal assemblage, the operators of successive
generations superposed and selectively canalised within one and the same cell.
Heterogeneity now holds in space and in time alike.

\subsection{Historicity}

It follows that the whole historical stratification remains available, and is
transmitted from generation to generation. Each assembly is a
laboratory for inventing operators. A modification arising in one generation,
is neither erased at the next division nor substituted for what
preceded it, but added to the repertoire that the offspring inherits and can
itself modify again. Successive inventions accumulate along the line of
descent, and the accumulated layers stay accessible rather than being
overwritten.

This is the specificity of biological objects that is historicity. A
biological object is not specified once and for all by a timeless law but is
defined by its history ~\cite{longo2014,longo2017}, understood as an organisation of events, that is by the
particular sequence of operator inventions that its ancestors assembled and transmitted, and by the
coefficients through which that sequence is currently canalised. The
assembly  \eqref{eq:system}  carry the accumulated repertoire in
the operators. In this sense the differential formulation reproduces, from
the reproduction condition alone, the historical character that
distinguishes living matter from the merely physical.

\section{The Gray--Scott kinetic model as an example}
\label{sec:gs}

The construction of Sections~\ref{sec:problem}--\ref{sec:historicity} leaves the
operators of the assemblage unspecified. We now look for concrete operators to
introduce into the assemblage such that
the resulting system satisfies the closure of constraints
(Condition~\ref{cond:fixed}) and the reproduction of the assembly
(Condition~\ref{cond:repro}). The Gray--Scott kinetics provides a natural
starting point, since in its homogeneous form it already exhibits a stable point of equilibrium and a close homoclinic orbit. 

\subsection{The Gray--Scott equations}
\label{sec:identical}
The Gray Scott model ~\cite{gray1984,pearson1993}  is a reaction-diffusion system of Turing type. 
Writing $\phi=(u,v)$ for the two concentrations, the Gray--Scott
model reads
\begin{equation}
\label{eq:gs}
\left\{
\begin{aligned}
\partial_t u &= D_u\,\Delta u - u v^{2} + F\,(1-u), \\[2pt]
\partial_t v &= D_v\,\Delta v + u v^{2} - (F+\kappa)\,v,
\end{aligned}
\right.
\end{equation}
where $D_u,D_v$ are diffusion coefficients, $F$ the feed rate and $\kappa$ the
removal rate. We write it compactly as $\phi_t=D\,\Delta\phi+R(\phi)$, with
$D=\mathrm{diag}(D_u,D_v)$ and $R$ the reaction term. For suitable parameters the
system supports localised spots, made of a raised plateau of $v$ over a depleted core of
$u$, with a sharp transition layer that plays the role of the interface
$\Gamma$.

\begin{figure}[htbp]
\centering
\includegraphics[width=\textwidth]{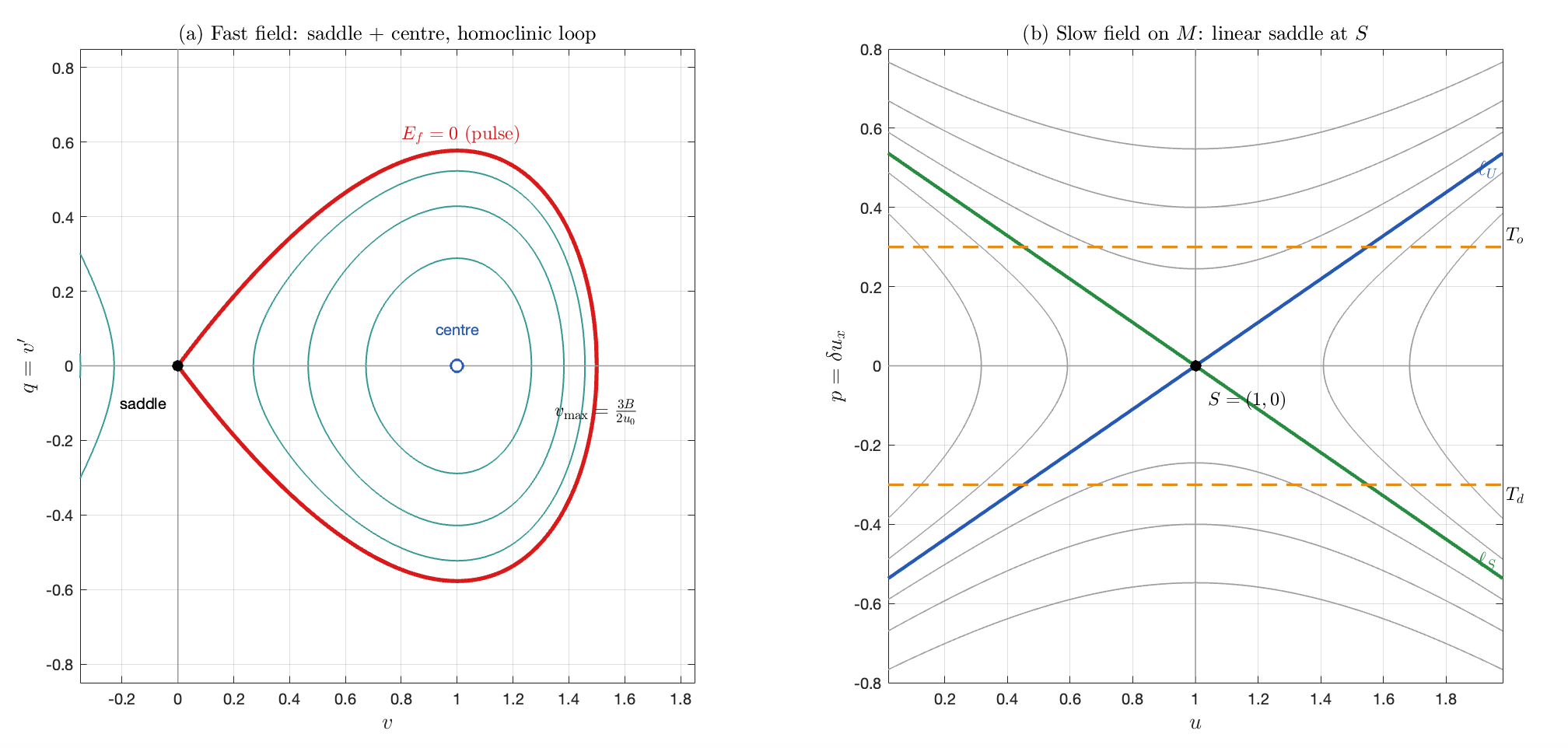}
\caption{The two reduced fields of the Gray--Scott profile equations.
(a) Fast field $(v,q)$ of \eqref{eq:fast-portrait}, with a saddle at the origin and a
centre at $(B/u_0,0)$; the thick red curve is the homoclinic level $E_f=0$ ---
the $\operatorname{sech}^{2}$ spot of peak $v_{\max}=3B/(2u_0)$ --- and the thin
curves are periodic levels $E_f<0$ inside the loop ($B=u_0=1$). This homoclinic
loop is the leaf-shaped orbit revolved in Figure~\ref{fig:cycle}.
(b) Slow field $(u,p)$ of \eqref{eq:slow-portrait}, with a linear saddle at
$S=(1,0)$ with eigenvalues $\pm\sqrt{A}$, unstable manifold $\ell_U$ and stable
manifold $\ell_S$ of \eqref{eq:invlines}, grey hyperbolic orbits
$p^{2}-A(u-1)^{2}=\mathrm{const}$, and the dashed take-off and touch-down lines
$T_o,T_d$ fixed by the jump condition \eqref{eq:melnikov} ($A=0.30$; the
$T_o,T_d$ offset drawn at a representative value). A one-spot solution exists
when the slow orbit out of $S$ along $\ell_U$, jumping from $T_o$ to $T_d$,
returns along $\ell_S$ and closes.}
\label{fig:gs-phase}
\end{figure}

\paragraph{Stability.} Following the geometric singular-perturbation analysis of
Doelman, Kaper and Zegeling~\cite{DKZ1997} and the stability analysis of
Doelman, Gardner and Kaper~\cite{DoelmanGardnerKaper1998}, the stationary
profile equations split, for a small ratio $\delta$ of the diffusion lengths,
into two reduced fields, both of saddle type. The \emph{fast field}
\begin{equation}
\label{eq:fast-portrait}
v''=B v-u_0 v^{2},\qquad
E_f(v,q)=\tfrac12 q^{2}-\tfrac12 B v^{2}+\tfrac13 u_0 v^{3},\qquad q=v',
\end{equation}
has a saddle at the origin and a centre at $(B/u_0,0)$; its level set $E_f=0$ is
the orbit homoclinic to the origin, which is the spot itself,
$v_0(x)=\tfrac{3B}{2u_0}\operatorname{sech}^{2}\!\bigl(\tfrac{\sqrt B}{2}x\bigr)$.
The \emph{slow field} on $M=\{v=q=0\}$,
\begin{equation}
\label{eq:slow-portrait}
\frac{du}{dX}=p,\qquad \frac{dp}{dX}=A\,(u-1),
\end{equation}
is a linear saddle at $S=(1,0)$ with eigenvalues $\pm\sqrt A$ and invariant lines
\begin{equation}
\label{eq:invlines}
\ell_U:\ p=+\sqrt{A}\,(u-1),\qquad \ell_S:\ p=-\sqrt{A}\,(u-1).
\end{equation}
A one-spot solution exists when the slow orbit leaving $S$ along $\ell_U$,
jumping across the fast loop between the take-off and touch-down lines fixed by
the Melnikov condition
\begin{equation}
\label{eq:melnikov}
\Delta p=\int_{-\infty}^{+\infty} u\,v_0(x)^{2}\,dx=O(\sqrt{B}),
\end{equation}
returns along $\ell_S$ and closes (Figure~\ref{fig:gs-phase}).

These two stability characteristics will facilitate the fulfillment of the closure and reproduction conditions, although they are not sufficient because individuation and assembly through the constraint $\Gamma$ are still lacking.
In facts when the one-spot homoclinic orbit
is stable, the profile persists and the field is a fixed point, while 
when it loses stability, it reorganises into
a two-spot orbit~\cite{ReynoldsPonceDawesProctor,MuratovOsipov}, in such a way that in the
$(\phi,\nabla\phi)$ plane one loop pinches into two, giving the closed orbit.

\subsection{The Gray-Scott assemblage}
\label{sec:hetero}

Finally we construct here the heterogenetic assemblage introducing the G-S kinetic equations in the system eqref{eq:system}. We heterogenise the operator across the interface in such a way that the exterior keeps the
Gray-Scott operator, while the interior replaces the constant diffusion by a
divergence-form operator whose conductivity carries the operator time $\tau$,
\begin{equation}
\label{eq:aniso}
\left\{
\begin{aligned}
\phi_t &= \operatorname{div}\bigl(g^{\tau}\,\nabla\phi\bigr)+R(\phi), && \phi(t,x) \ge \phibar, \\[2pt]
\phi_t &= D\,\Delta\phi+R(\phi), && \phi(t,x) < \phibar, \\[2pt]
\Gamma(t) &= \bigl\{\, x : \phi(t,x) = \phibar \,\bigr\}, && \\[2pt]
\phi(0,x) &= \phi_0(x), && x \in \Omega,
\end{aligned}
\right.
\end{equation}
where the conductivity is a function of the direction of the gradient,
\[
g^{\tau}=g^{\tau}\!\Bigl(\frac{\nabla u}{|\nabla u|}\Bigr).
\]
The operator thus reads the local orientation of the interface and diffuses
anisotropically with respect to it. It is not a datum attached to the points of
the domain but a function of the solution's own gradient, and the superscript
$\tau$ records that it may vary along the history of the operators. Note that
$\operatorname{div}(g^{\tau}\nabla\phi)$ is the entire diffusive term and not
a diffusion coefficient. Expanding the divergence separates it into a parabolic
part and two advective parts,
\begin{equation}
\label{eq:aniso-expanded}
\operatorname{div}\bigl(g\,\nabla\phi\bigr)=g\,\Delta\phi+g_x\,\phi_x+g_y\,\phi_y,
\end{equation}
a separation that dictates the discretisation of Section~\ref{sec:numerics}.
Different choices of $g^{\tau}$ produce different interfacial geometries, round, hexagonal, square, rotating or asymmetric, and the
sequence of these
choices along $\tau$ is the temporal assemblage of Section~\ref{sec:historicity}.

For this heterogeneous system we cannot at present establish analytically that
the two conditions hold, because the coupling of two distinct operators across a moving
interface is beyond the reach of the arguments available in the homogeneous
case. Numerical simulation shows, however, that both are fulfilled, as
documented in Section~\ref{sec:numerics}.

\section{Numerical simulations}
\label{sec:numerics}

\subsection{The discretised domain}

All simulations are carried out on the unit square
\[
\Omega=[0,1]\times[0,1]\subset\Rtwo,
\]
discretised by a uniform Cartesian lattice of spacing $\Delta x$ in both
directions,
\[
x_j=j\,\Delta x,\qquad y_k=k\,\Delta x,\qquad j,k=0,\dots,N,\qquad N\Delta x=1,
\]
so that a field is represented by its nodal values
$u_{j,k}^{\,m}\simeq u(x_j,y_k,t_m)$, with $t_m=m\,\Delta t$. Periodic boundary
conditions are imposed on $\partial\Omega$, so that the domain is a flat torus
and no artificial wall interferes with the free boundary $\Gamma$.

\subsection{Approximation of operators}

The operators $\Op_i$ and $\Op_o$ are approximated by finite differences in
space and in time. The type of each term dictates its stencil. Parabolic (diffusive) terms are discretised with
\emph{centred} differences, which are second-order accurate and preserve the
symmetry of the Laplacian, advective (transport) terms are discretised with
\emph{upwind} differences, which respect the direction of propagation and keep
the scheme monotone, and the evolution in time is advanced by \emph{forward}
differences.

\paragraph{Parabolic terms: centred differences.}
The Laplacian is replaced by the standard five-point stencil,
\begin{equation}
\label{eq:lap5}
\bigl(\Delta_h u\bigr)_{j,k}
=\frac{u_{j+1,k}+u_{j-1,k}+u_{j,k+1}+u_{j,k-1}-4\,u_{j,k}}{\Delta x^{2}}
=\Delta u(x_j,y_k)+O(\Delta x^{2}),
\end{equation}
and first derivatives appearing inside a diffusive flux are likewise centred,
\begin{equation}
\label{eq:centred}
\bigl(\delta^{c}_{x}u\bigr)_{j,k}=\frac{u_{j+1,k}-u_{j-1,k}}{2\,\Delta x},
\qquad
\bigl(\delta^{c}_{y}u\bigr)_{j,k}=\frac{u_{j,k+1}-u_{j,k-1}}{2\,\Delta x}.
\end{equation}

\paragraph{Advective terms: upwind differences.}
A term of the form $a\,\partial_x u$ is discretised according to the sign of the
local velocity $a=a_{j,k}$,
\begin{equation}
\label{eq:upwind}
\bigl(a\,\delta^{u}_{x}u\bigr)_{j,k}=
a_{j,k}^{+}\,\frac{u_{j,k}-u_{j-1,k}}{\Delta x}
+a_{j,k}^{-}\,\frac{u_{j+1,k}-u_{j,k}}{\Delta x},
\qquad
a^{\pm}=\tfrac12\bigl(a\pm|a|\bigr),
\end{equation}
and analogously in $y$. Information is thus always taken from the side from
which it arrives; the scheme is first-order accurate but free of the spurious
oscillations that a centred stencil produces on a transport term.

\paragraph{Time stepping: forward differences.}
Writing $\mathcal{L}_h$ for the full spatial discretisation of the right-hand
side, the fields are advanced explicitly,
\begin{equation}
\label{eq:euler}
\frac{u^{\,m+1}_{j,k}-u^{\,m}_{j,k}}{\Delta t}
=\bigl(\mathcal{L}_h u^{\,m},\,v^{\,m}\bigr)_{j,k},
\end{equation}
which is the forward (explicit) Euler scheme. Stability of the diffusive part
requires the parabolic restriction
\begin{equation}
\label{eq:cfl}
\Delta t\;\le\;\frac{\Delta x^{2}}{4\,\max\{D_u,D_v\}},
\end{equation}
and the advective part the Courant condition
$\Delta t\,\max|a|\le\Delta x$, and the time step is chosen as the smaller of the
two. The interface $\Gamma$ is not tracked explicitly, and at each step it is
recovered as the discrete level set $\{\phi_{j,k}=\phibar\}$, so that the
free-boundary coupling of \eqref{eq:system} is enforced automatically by the
level-set representation.

\subsection{Homogeneous case}

\begin{figure}[htbp]
\centering
\includegraphics[width=0.245\textwidth]{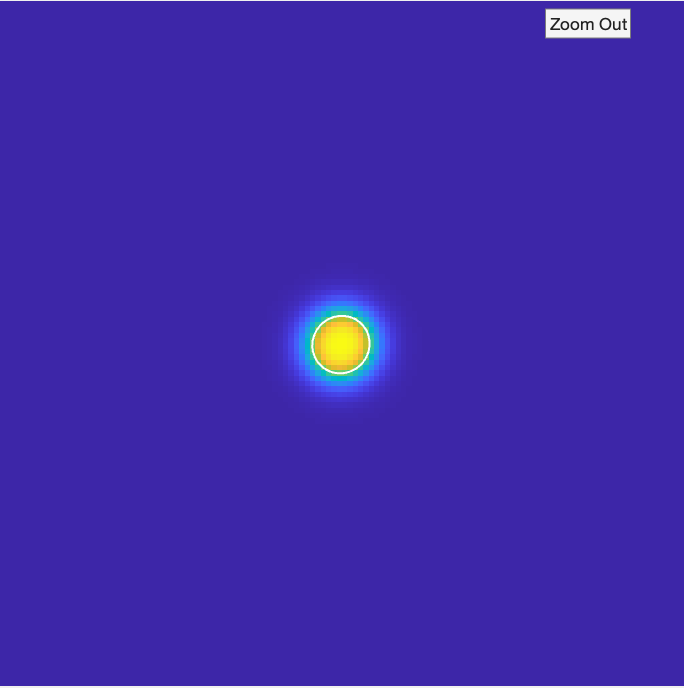}\hfill
\includegraphics[width=0.245\textwidth]{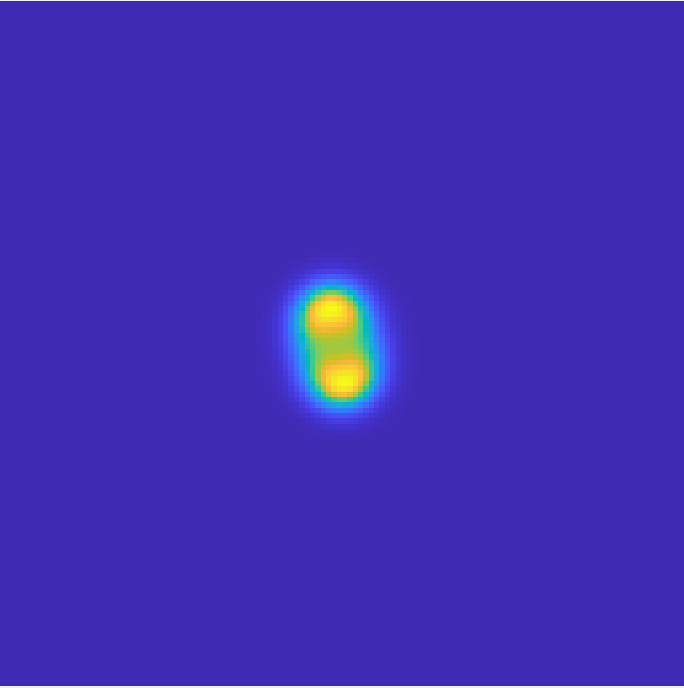}\hfill
\includegraphics[width=0.245\textwidth]{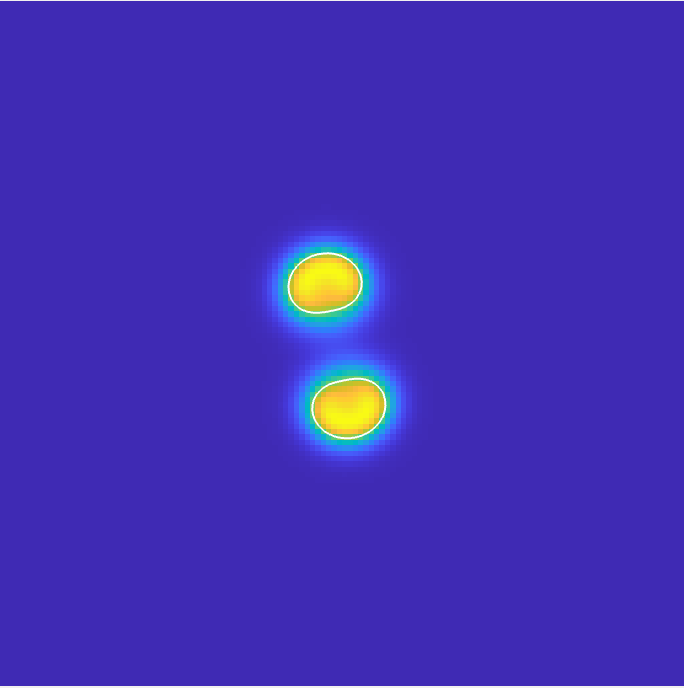}\hfill
\includegraphics[width=0.245\textwidth]{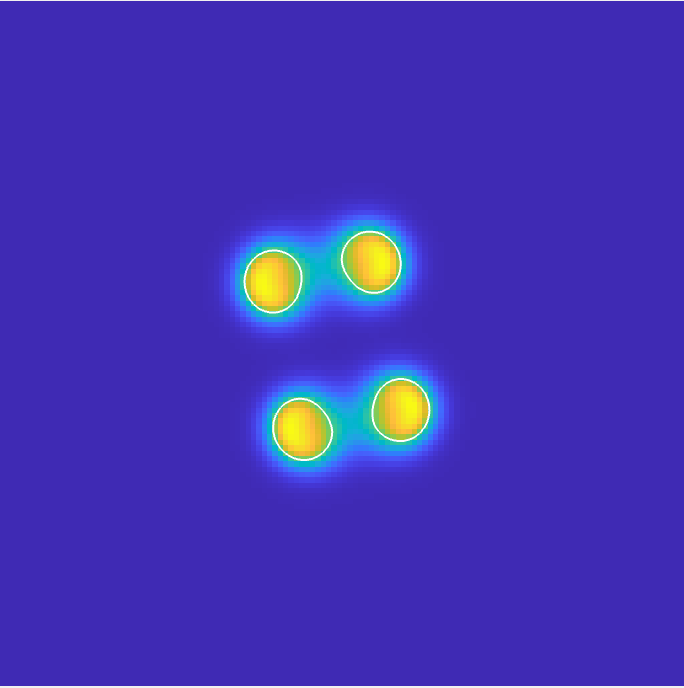}\\[3pt]
\includegraphics[width=0.245\textwidth]{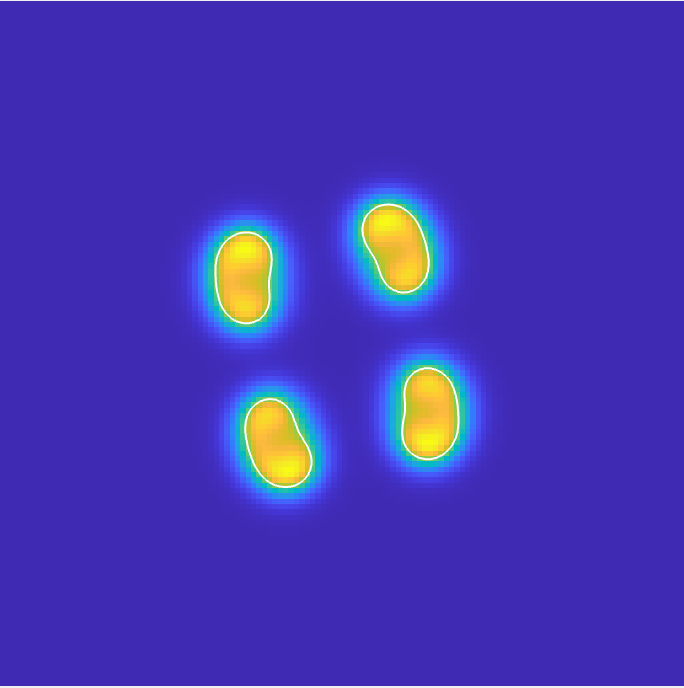}\hfill
\includegraphics[width=0.245\textwidth]{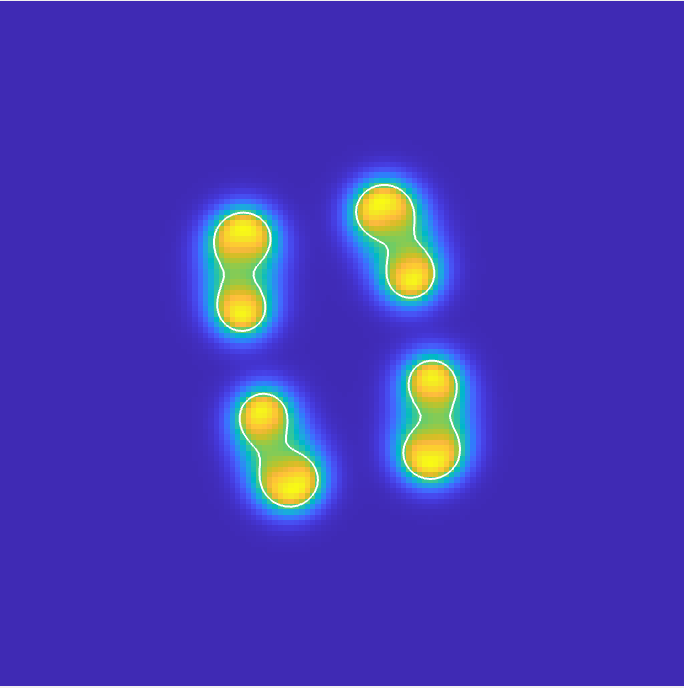}\hfill
\includegraphics[width=0.245\textwidth]{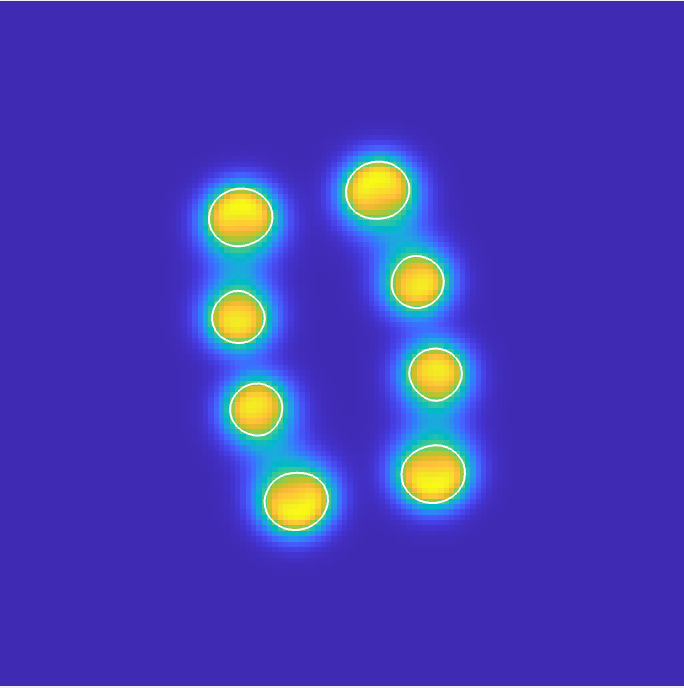}\hfill
\includegraphics[width=0.245\textwidth]{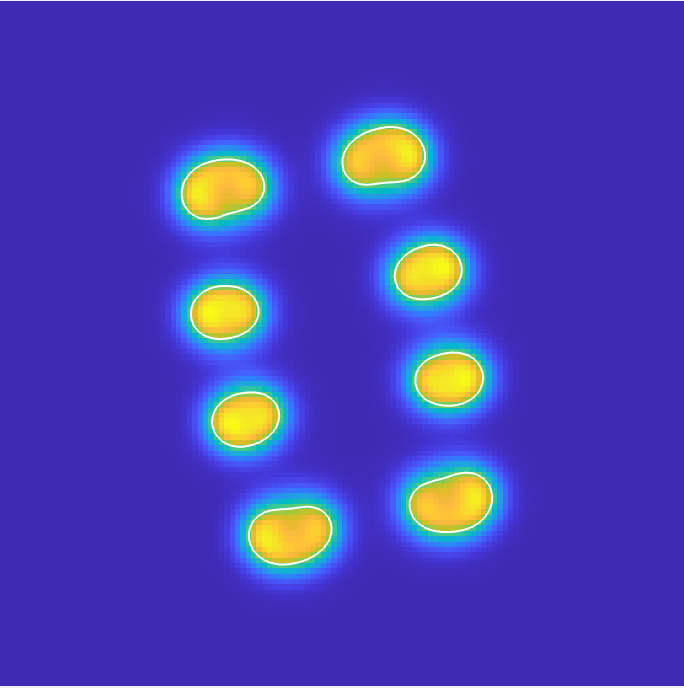}\\[3pt]
\includegraphics[width=0.245\textwidth]{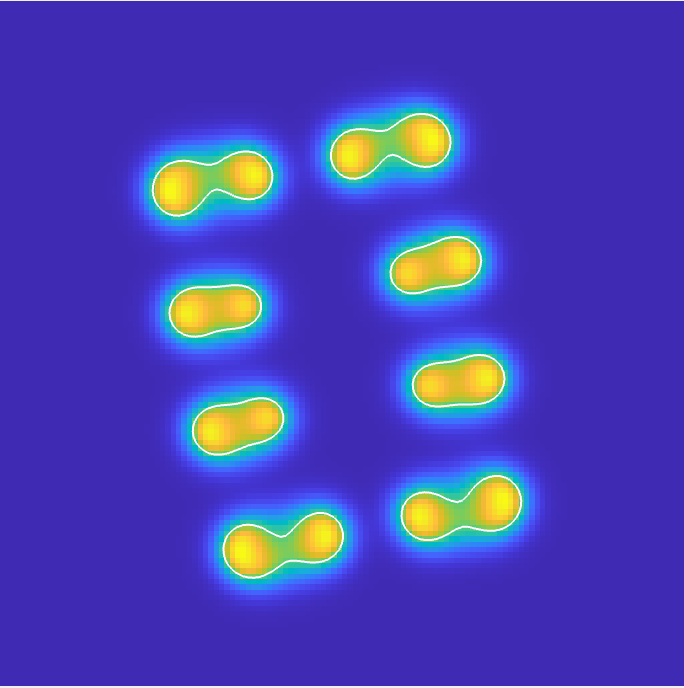}\hfill
\includegraphics[width=0.245\textwidth]{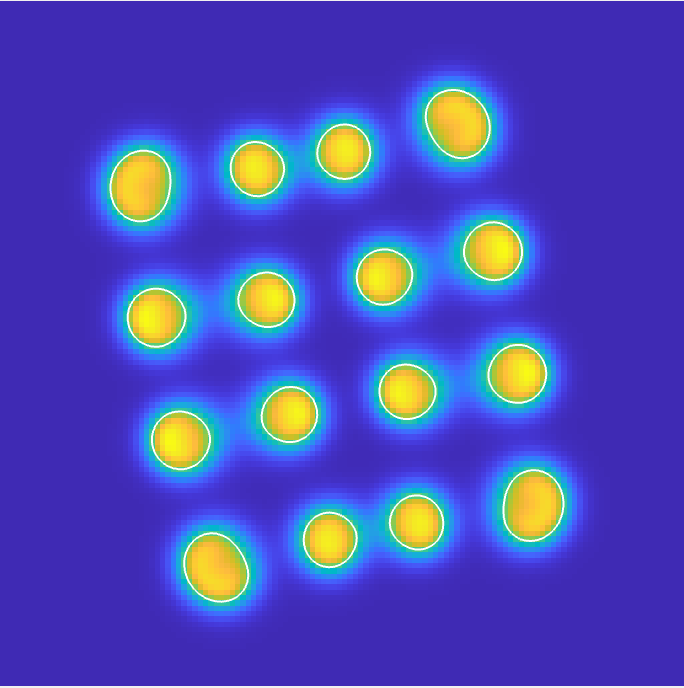}\hfill
\includegraphics[width=0.245\textwidth]{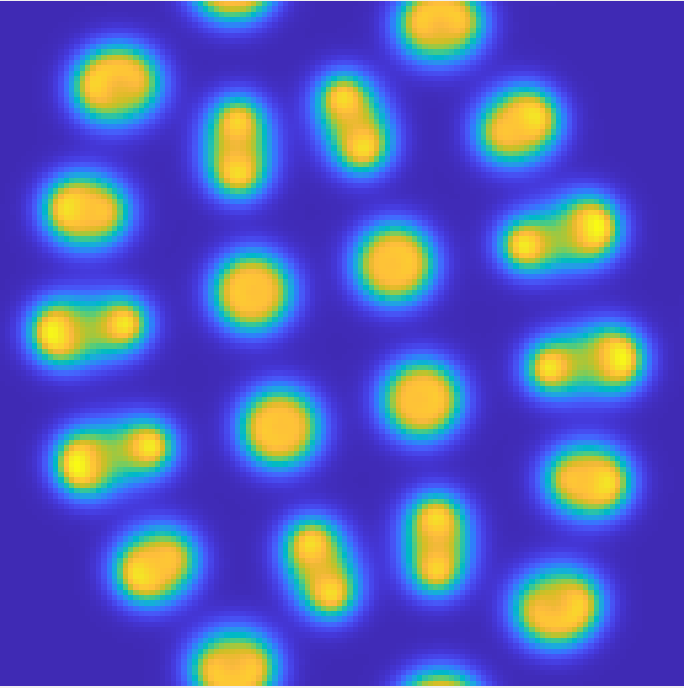}\hfill
\includegraphics[width=0.245\textwidth]{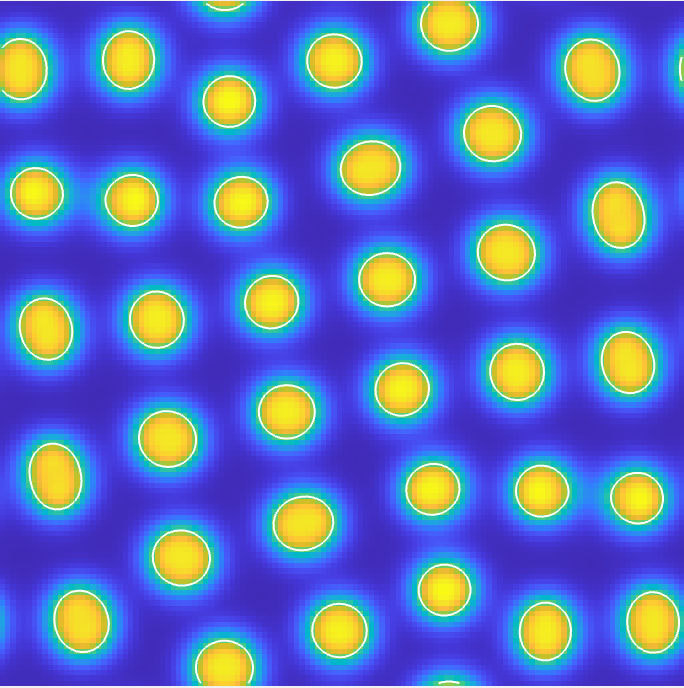}
\caption{Homogeneous Gray-Scott dynamics ($\Op_i=\Op_o$) on
$[0,1]^{2}$, twelve successive instants read left to right and top to bottom.
Starting from a single mother cell, the spot duplicates, the daughters
duplicate in turn, and the multiplication proceeds until the domain is filled.
Each division is one completion of the cycle $C_{\Ass}^{\tau}$.}
\label{fig:num-gs}
\end{figure}

The first series of simulations integrates the \emph{homogeneous} discretised
Gray-Scott system, i.e.\ the case $\Op_i=\Op_o$ of Section~\ref{sec:identical}, in the
self-replicating parameter band. The initial datum is a single localised cell.
Figure~\ref{fig:num-gs} shows twelve successive instants, in which the mother cell
elongates and divides, each daughter divides in turn, and the progeny fill the
domain until the available substrate is exhausted and the population saturates.
This is the repeated action of the cycle $C_{\Ass}^{\tau}$ of
Condition~\ref{cond:repro}, iterated over generations.

\subsection{Heterogeneous case}

The second series realises the heterogeneous assembly introduced in
Section~\ref{sec:hetero}, in which the exterior operator $\Op_o$ is the usual
Gray--Scott right-hand side, while the interior operator $\Op_i$ carries the
anisotropic, gradient-dependent diffusion of \eqref{eq:aniso}. Its expansion
\eqref{eq:aniso-expanded} dictates the discretisation. The
parabolic terms $g\,\Delta u$ and $g\,\Delta v$ are treated with the centred
stencil \eqref{eq:lap5}, while the \emph{advective} terms $g_x u_x+g_y u_y$ and
$g_x v_x+g_y v_y$ are treated with the upwind scheme \eqref{eq:upwind}, the
local velocity being $a=(g_x,g_y)$. A centred stencil on the
advective part produces oscillations at the interface and destroys the cell
shape; the upwind choice is what makes the heterogeneous simulations robust.

\begin{figure}[htbp]
\centering
\includegraphics[width=0.32\textwidth]{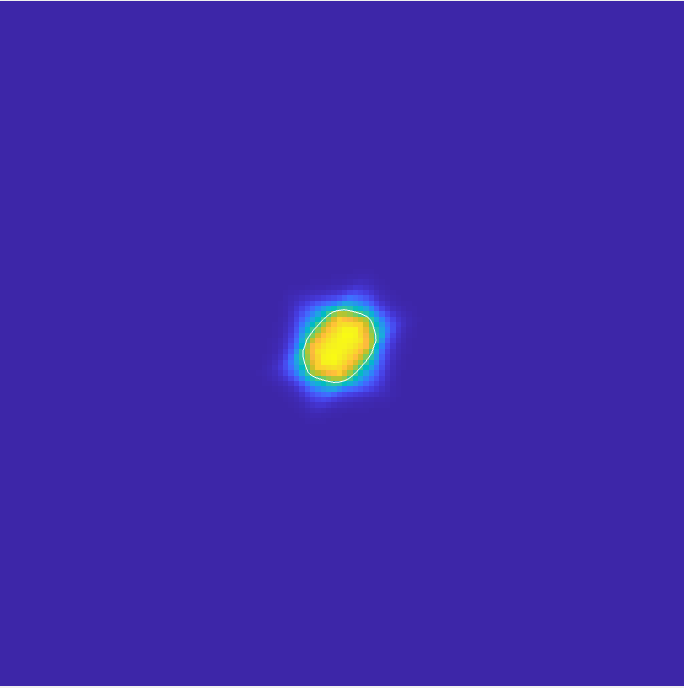}\hfill
\includegraphics[width=0.32\textwidth]{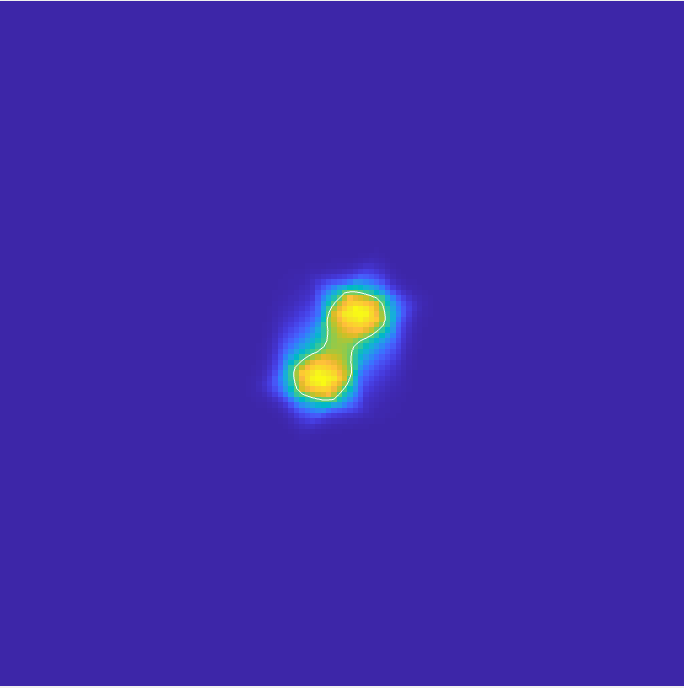}\hfill
\includegraphics[width=0.32\textwidth]{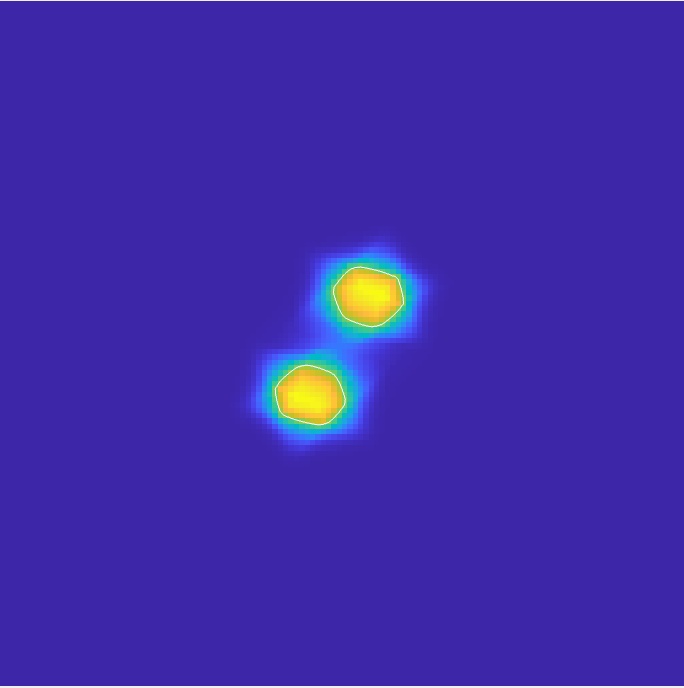}\\[3pt]
\includegraphics[width=0.32\textwidth]{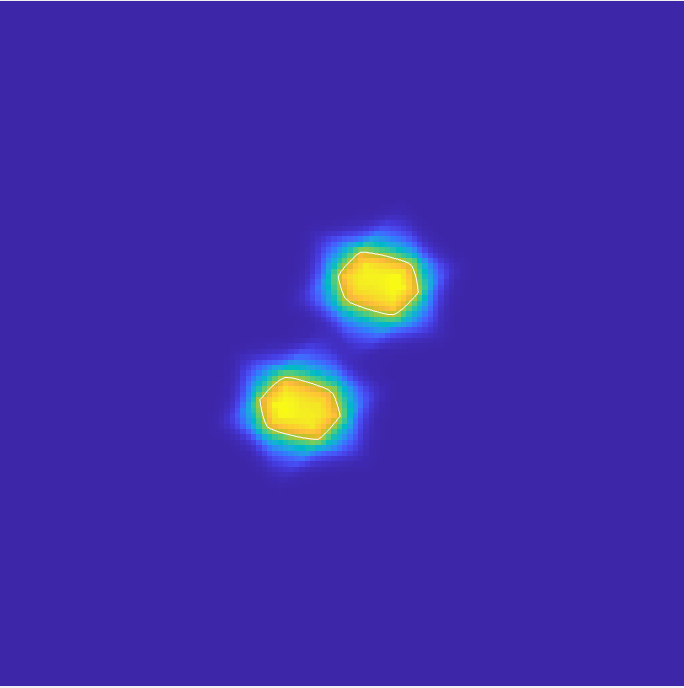}\hfill
\includegraphics[width=0.32\textwidth]{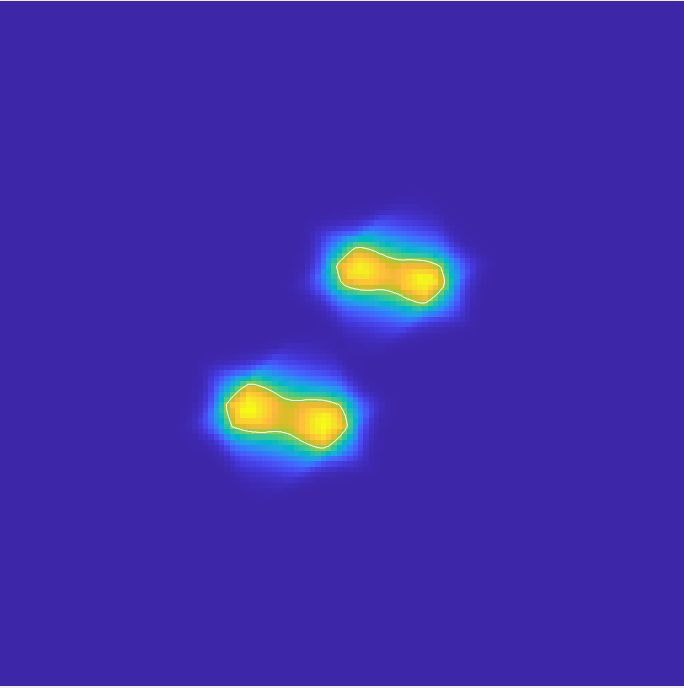}\hfill
\includegraphics[width=0.32\textwidth]{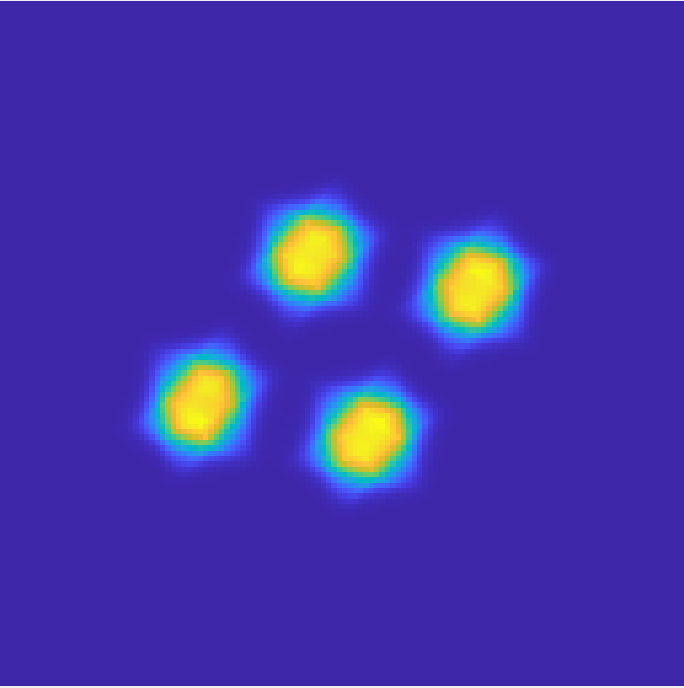}
\caption{Heterogeneous assembly with anisotropic interior diffusion
\eqref{eq:aniso} and conductivity \eqref{eq:g-hex}. The mother cell and all its
progeny are hexagonal, since the operator change alters the geometry of $\Gamma$
while the reproduction condition continues to hold, so the new shape is
transmitted at every division.}
\label{fig:num-hex}
\end{figure}

\paragraph{Hexagonal cells.}
Taking the conductivity to depend on the direction of the gradient,
\begin{equation}
\label{eq:g-hex}
g=\arctan\!\Bigl(6\,\frac{\nabla u}{|\nabla u|}\Bigr),
\end{equation}
breaks the rotational symmetry of the interface and selects a preferred set of
angular directions. Figure~\ref{fig:num-hex} shows the resulting time series:
starting again from a single mother cell, the progeny are no longer round but
hexagonal, and they multiply while preserving that shape.

\begin{figure}[htbp]
\centering
\includegraphics[width=0.32\textwidth]{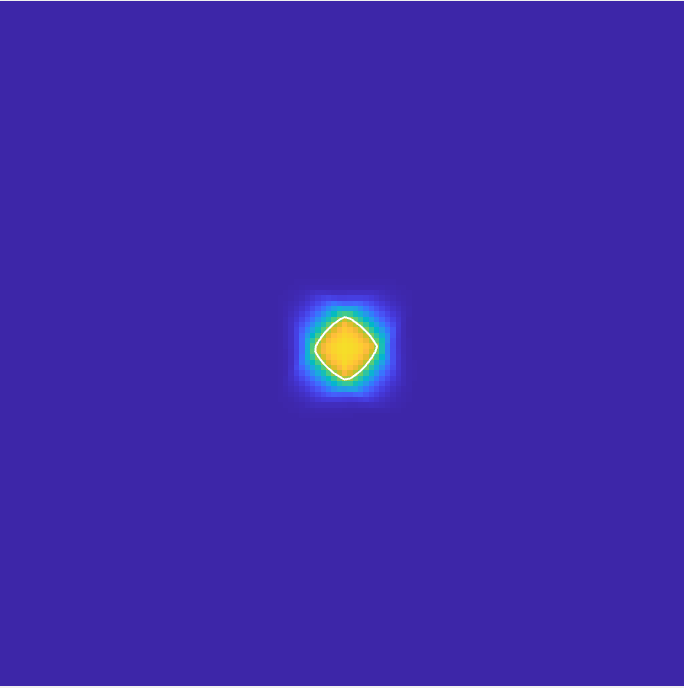}\hfill
\includegraphics[width=0.32\textwidth]{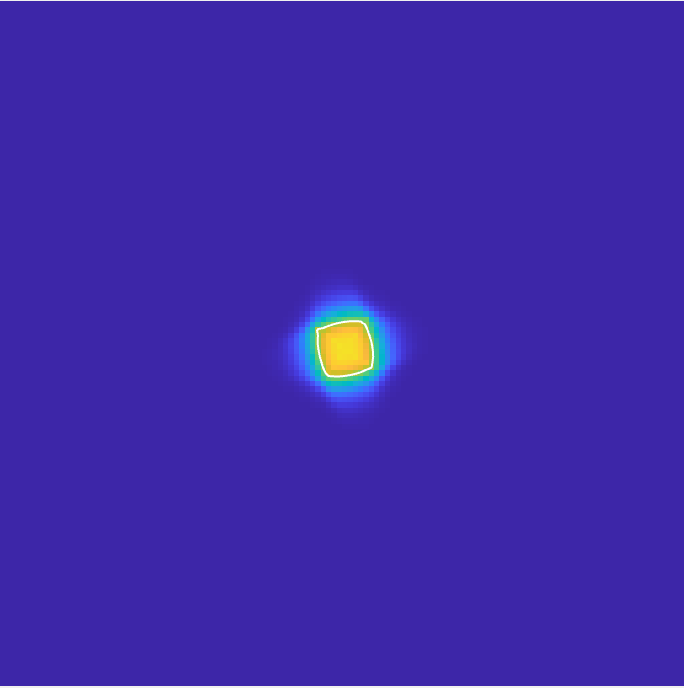}\hfill
\includegraphics[width=0.32\textwidth]{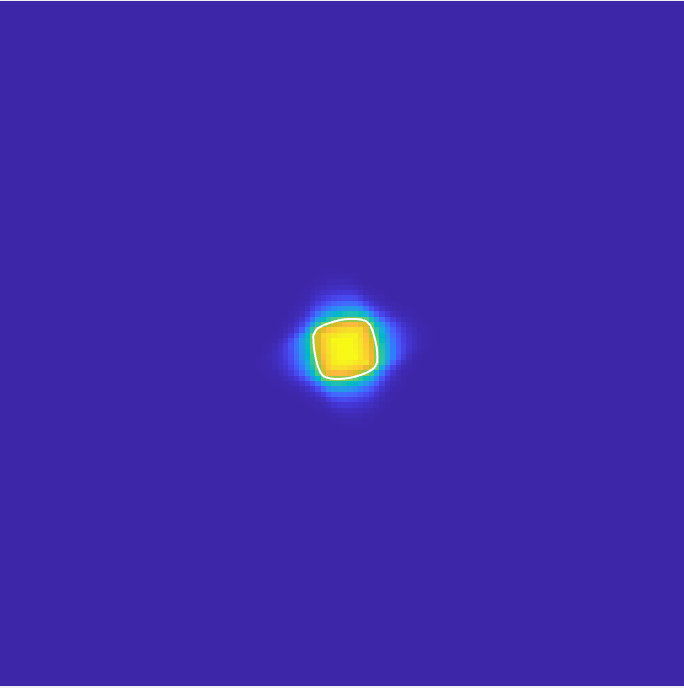}\\[3pt]
\includegraphics[width=0.32\textwidth]{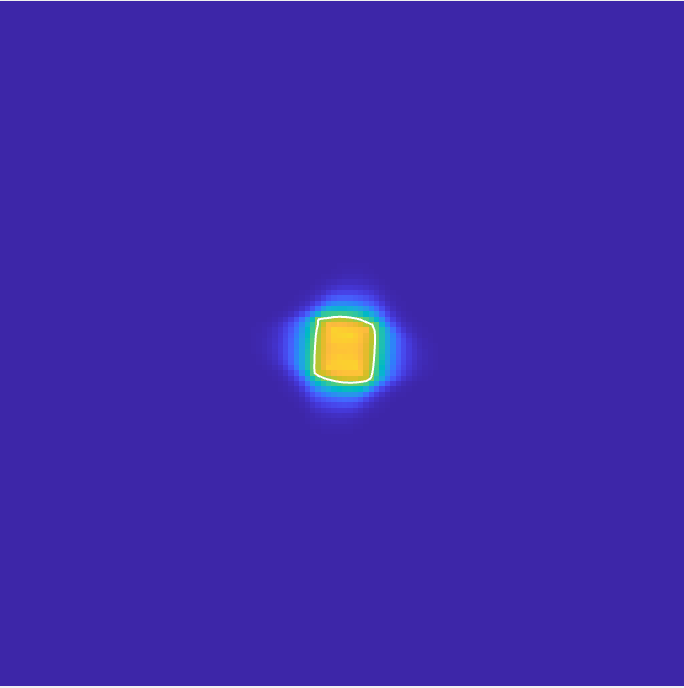}\hfill
\includegraphics[width=0.32\textwidth]{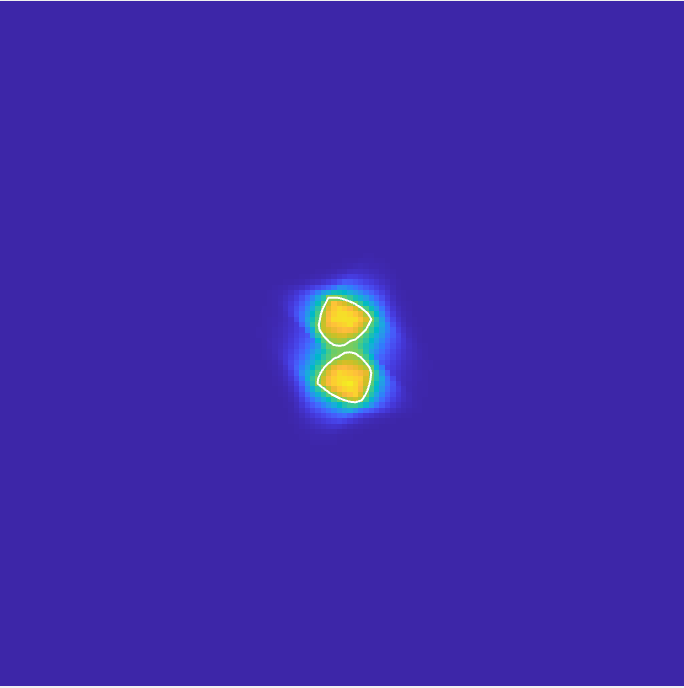}\hfill
\includegraphics[width=0.32\textwidth]{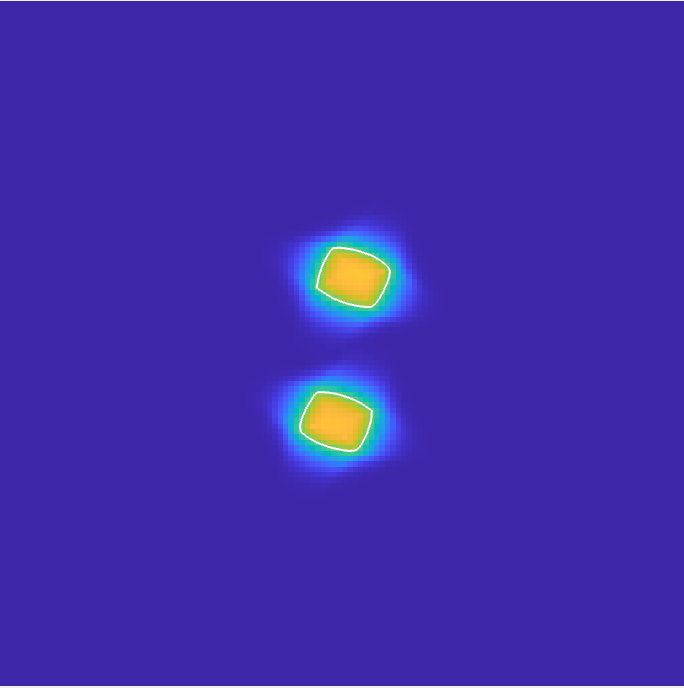}\\[3pt]
\includegraphics[width=0.32\textwidth]{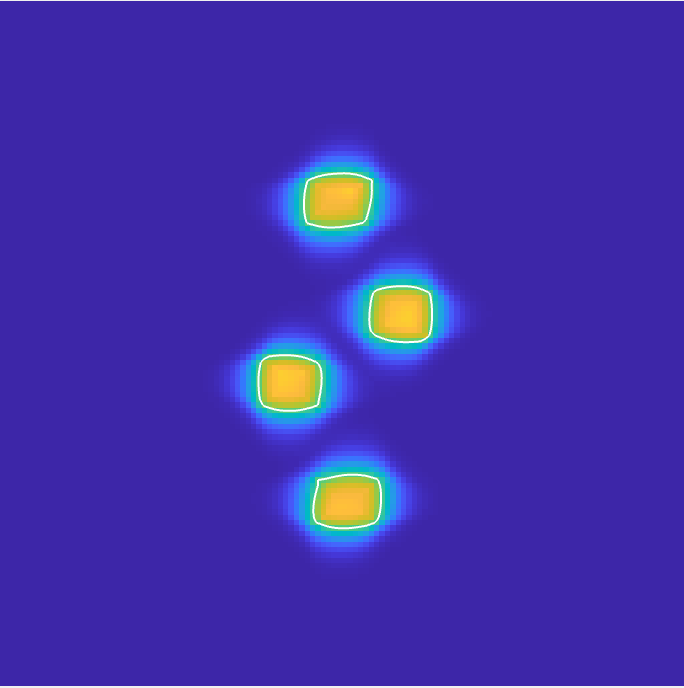}\hfill
\includegraphics[width=0.32\textwidth]{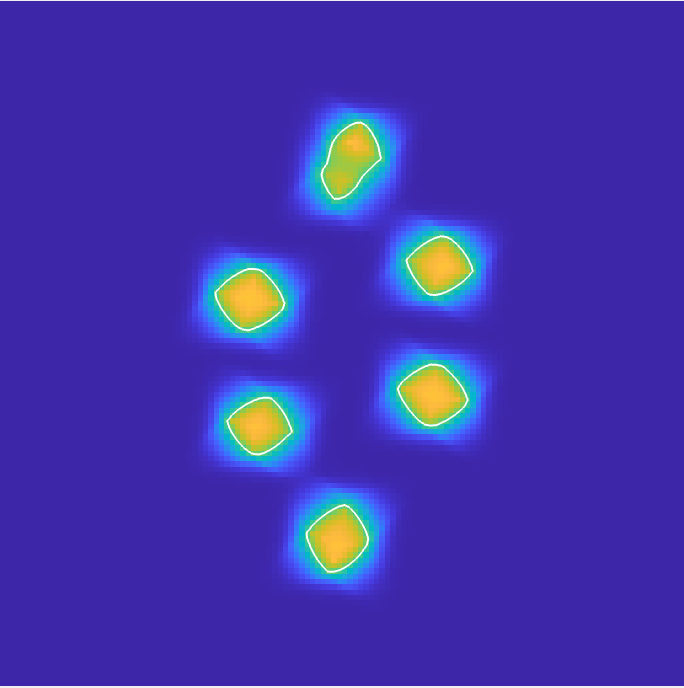}\hfill
\includegraphics[width=0.32\textwidth]{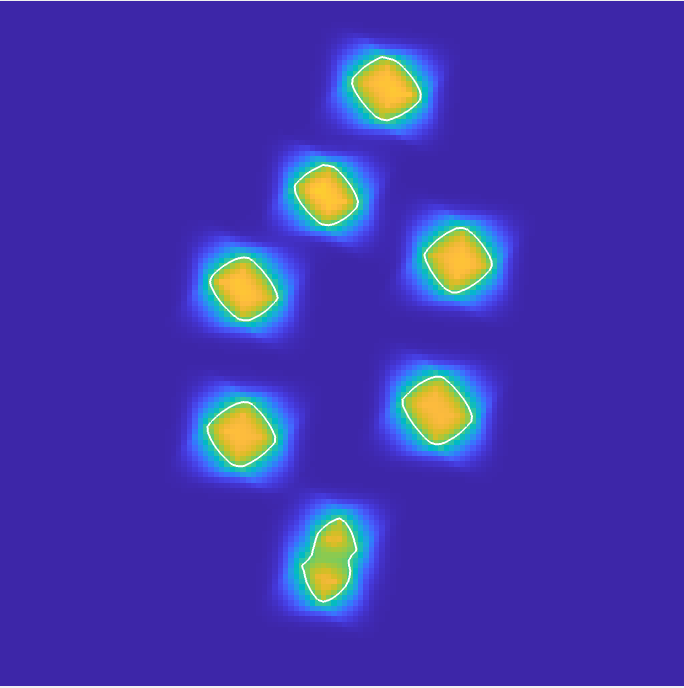}
\caption{Rotating square cells obtained with the time-dependent conductivity
\eqref{eq:g-rot}, nine successive instants. The interface selects four
preferred directions and the resulting square precesses in time.}
\label{fig:num-rot}
\end{figure}

\paragraph{Rotating square cells.}
Letting the same construction depend explicitly on time,
\begin{equation}
\label{eq:g-rot}
g=\arctan\!\Bigl(4\,\frac{\nabla u}{|\nabla u|}+t\Bigr),
\end{equation}
makes the preferred directions precess. The cells are now square, and
the square slowly rotates as the simulation proceeds
(Figure~\ref{fig:num-rot}). The assembly is still reproduced, but the
reproduced object now carries an internal orientation that evolves, a
further degree of freedom created by the choice of operator.

\subsection{Variation}

The last simulations show the change of operator discussed in
Section~\ref{sec:historicity} within a running population. Here the operators
are no longer kept fixed and are allowed to vary along
$\tau$, so that the interior operator becomes a function of the operator time.
The variation is written as a switch at a sequence of instants $\tau_k$,
\begin{equation}
\label{eq:g-variation}
g^{\tau}=g+\mathcal{H}(\tau-\tau_k)\,\delta g,
\end{equation}
where $\mathcal{H}$ is the Heaviside function and $\delta g$ the increment
introduced at the $k$-th event in the time $\tau$. Before $\tau_k$ the cell is governed by $g$;
after it, by $g+\delta g$. The switch is instantaneous in $\tau$, and once made
it is not undone. The new conductivity is what the cell carries from then on,
and what it transmits at division.

\begin{figure}[htbp]
\centering
\begin{minipage}{0.48\textwidth}\centering
\includegraphics[width=\textwidth]{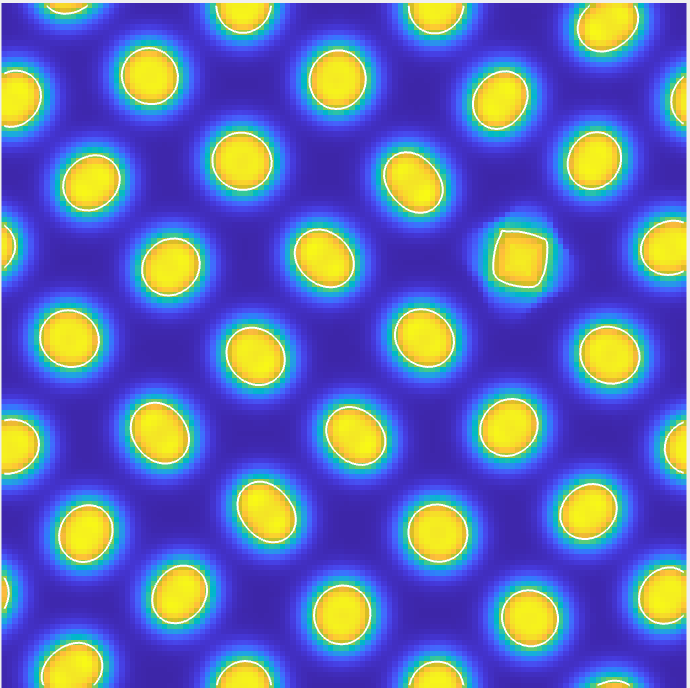}\\[2pt]
{\footnotesize (a) one cell varies}
\end{minipage}\hfill
\begin{minipage}{0.48\textwidth}\centering
\includegraphics[width=\textwidth]{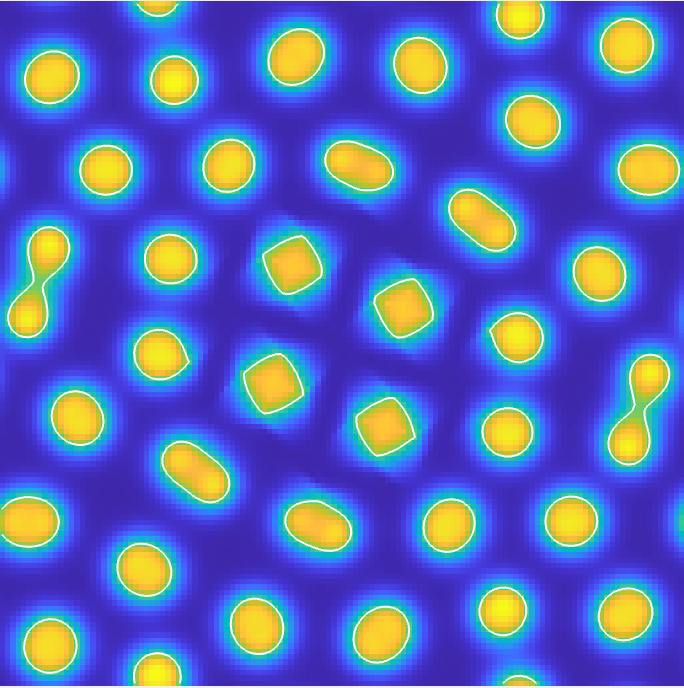}\\[2pt]
{\footnotesize (b) four cells vary}
\end{minipage}
\caption{Variation of the operators. (a) A single Gray--Scott cell becomes a
square cell when its interior operator is switched by \eqref{eq:g-variation}.
(b) A group of four Gray--Scott cells varies into square cells. The acquired
operator, and hence the acquired shape, persists in the offspring. It should be
noted that as soon as single cells, or groups of cells, vary independently, the
domain carries more than two phases, each varied lineage being is a phase of
its own, governed by its own operator. We used the classical technique of Rosenfeld and Pfaltz~\cite{R-P1966} to label and track internal domains $\Omega_i^j$}
\label{fig:num-mut}
\end{figure}

In Figure~\ref{fig:num-mut}(a) a single Gray--Scott cell is switched to the
anisotropic operator and becomes a square cell; in Figure~\ref{fig:num-mut}(b)
a group of four Gray--Scott cells undergoes the same variation together. In both
cases the varied geometry is not transient. It is retained by the cell and,
through the reproduction condition, handed on to its descendants. The population
thereafter contains a lineage defined by the operator its ancestor acquired, which is the elementary form of historicity.

\subsection{Reproduction with variation}

\begin{figure}[htbp]
\centering
\includegraphics[width=0.98\textwidth]{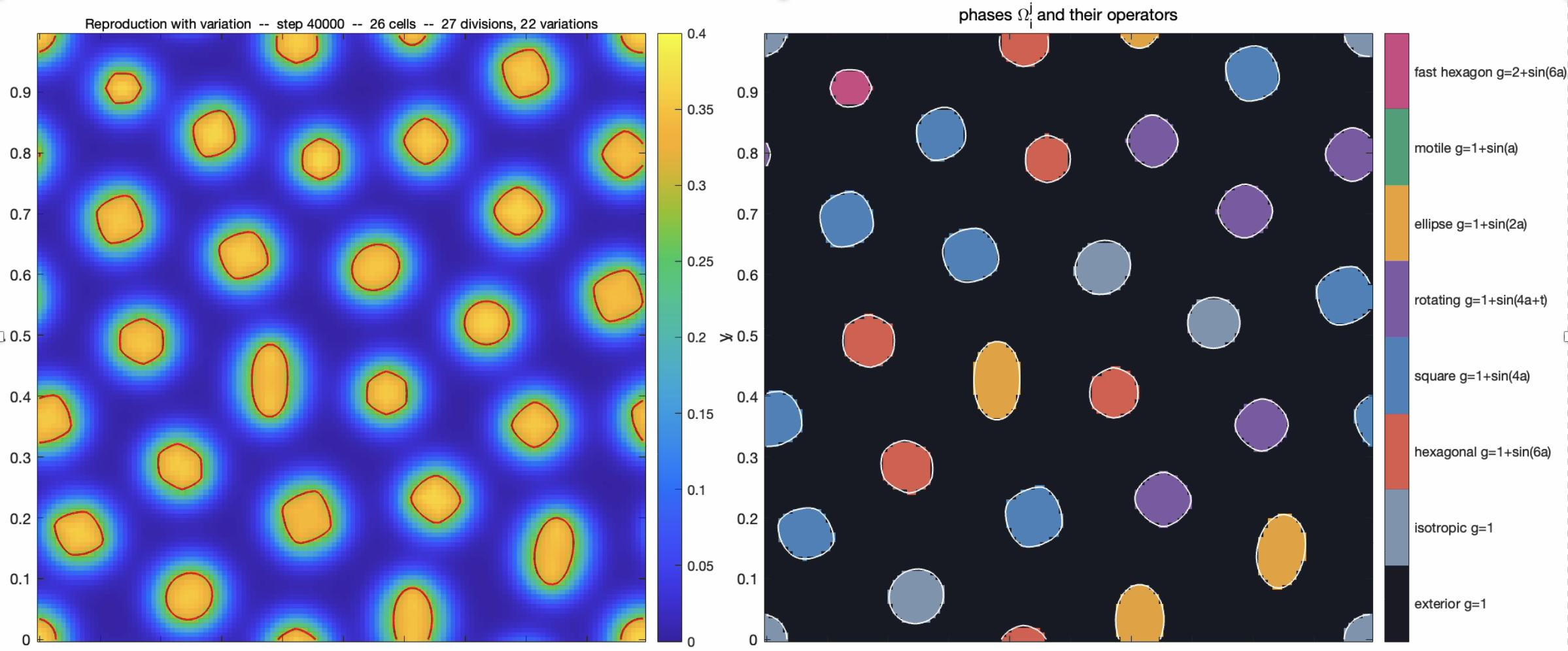}
\caption{Reproduction with variation. On the left, the population obtained by varying
the operator by $\delta g$ at successive divisions, each daughter inheriting the
operator of its mother unless the Heaviside switch \eqref{eq:g-variation} fires.
The cells are round, hexagonal, square and elliptic, and coexist in the same
domain, forming a ``fruit salad'' assembled by reproduction rather than imposed at
the initial datum. On the right, the corresponding phase map, each colour marking one
of the operators $g^{\tau}$, so that the lineages defined by their operators are
read directly. Because each lineage is governed by its own operator, the domain
carries more than two phases, which is why the interface $\Gamma$ is no longer a
single curve but a family of disjoint boundaries.}
\label{fig:num-macedonia}
\end{figure}

The final simulation illustrates the purpose of the construction. A single Gray--Scott
cell is left to reproduce, and at each division the operator of one daughter is
varied according to \eqref{eq:g-variation}, each with a different increment
$\delta g$ taken from the menu of Section~\ref{sec:hetero}. Nothing is imposed
on the population as a whole, and the shape of each cell is decided locally by the
operator it inherited, and the diversity visible in
Figure~\ref{fig:num-macedonia} is produced generation by generation. 
Note that when a multiphase problem involves more than two phases, as in this case, it is necessary to use a technique to label and track the phases. We used the classical technique of Rosenfeld and Pfaltz~\cite{R-P1966} for the labelling of internal domains $\Omega_i^j$.

The shapes are not prefixed but are the outcome of the operators that happened to be transmitted, so the same initial
condition with a different sequence of switches gives a different population.
The coexistence of several operators is what the temporal
assemblage describes, each cell carrying its own stratification, and the
population as a whole exhibits the diversity that accumulation along $\tau$
produces.

\section{Discussion}
\label{sec:discussion}

\subsection{From differential heterogenesis to a theory of the living}
\label{sec:heterogenesis}

The theory proposed here is based on the idea of adopting the differential heterogenesis framework ~\cite{sarti2018,sarti2022} to develop a theory of living systems that can reconcile operational closure and historicity. 
In this setting the primitive object of a morphogenetic theory is not a single homogeneous law acting
on a domain but a composition of assemblages of differential operators changing over time. 
The present paper adds a pair of constraints to such assemblages: the fixed-point principle and the principle of reproduction with variation. It also claims that these two constraints transform a heterogenetic construction into a living construction.. In this way, heterogenesis provides generative freedom, and the two conditions allow the assemblage to persist and reproduce.

The move can be summarised in the scheme  given below.

\begin{table}[htbp]
\centering
\small
\renewcommand{\arraystretch}{1.35}
\begin{tabular}{@{}p{0.30\textwidth}p{0.30\textwidth}p{0.32\textwidth}@{}}
\hline
\textbf{Already given by differential heterogenesis} & \textbf{Constraint
added here} & \textbf{What the constraint yields}\\
\hline
Spatial assemblage of heterogeneous operators composed over a domain &
\emph{Fixed-point condition}
$\Ass(\phi)=\phi$, hence $\Ass(\Gamma)=\Gamma$ &
the assemblage persists as a solution of its own production, and the
form it generates is self-maintaining rather than transient\\
Temporal assemblage of operators superposed and modulated in time &
\emph{Reproduction condition}
$C_{\Ass}^{\tau}(\Ass)=\Ass$ &
the assemblage is reproduced with variation, since the whole stratification of
operators is transmitted, so the form generated is inherited rather than
re-invented\\
\hline

\end{tabular}
\caption{From heterogenesis to the living (Section~\ref{sec:heterogenesis}).
The differential assemblages are already given; what must be added are the two
structural constraints of Section~\ref{sec:conditions}. Heterogenesis
supplies the generative freedom --- any composition of operators may produce a
form. The two constraints supply the closure that selects, among all such
compositions, those that constitute an individual, one that maintains
itself and reproduces itself together with its own history. A general
morphogenetic theory thereby becomes an approach to living systems.}
\label{tab:scheme-heterogenesis}
\end{table}

\subsection{Three steps away from physico-chemical kinetics}
\label{sec:threesteps}

On the other way we could begin from ordinary physico-chemical kinetics, for example a reaction--diffusion system whose
equation is invariant in space and in time, and go towards heterogenesis with three successive steps:

The first step is to instantiate a process of individuation through the boundary $\Gamma=\{\phi=\phibar\}$. Introducing the interface
$\Gamma$ as a level set of the solution itself converts the
problem into a \emph{free-boundary} problem. The cell ceases to be a pattern
spread over an undifferentiated field and becomes individuated, something
delimited whose limit is not imposed from outside but defined from within by
the very dynamics it delimits. 

The second move is the spatial heterogeneous assemblage. Letting $\Op_i$ and
$\Op_o$ be distinct operators on the two sides of $\Gamma$ destroys
the spatial invariance of the original law. The system is no longer one
equation on a domain but two operators in \emph{spatial assemblage}, glued
along a boundary that each of them helps to move. Interior and exterior become
qualitatively different milieux rather than two regions of the same milieu.

The third move is the temporal heterogeneous assemblage. Allowing the operators
to vary, at duplication or during the life of the cell, and letting the old and the new operator coexist in the superposition
$\Op_i=\sum_k a_k(t)\,\Op_{ik}$ of \eqref{eq:time-assemblage-n}, destroys the
temporal invariance as well. The law is no longer given once and for all but is
itself a variable of the problem, assembled in time out of its own history.

The result of these three steps is that a physical system whose equation was
invariant in space and in time has been transformed into a differential system
that is \emph{heterogeneous}, \emph{free-boundary}, and \emph{only partially
autonomous}, autonomous in that the boundary and the field determine one
another, but only partially so, because the temporal variation of the operators
is not itself fixed by the equations, but it is left to an event-based dynamics. 
This residual non-autonomy is not a defect
of the formulation but the locus at which novelty can enter, and
therefore the locus at which the model stops being a physical model and starts
being a biological one. This partial autonomy allows the theory to be open to interior and exterior events.

This path can be summarised in a scheme dual to
Table~\ref{tab:scheme-heterogenesis}. 
\begin{table}[htbp]
\centering
\small
\renewcommand{\arraystretch}{1.35}
\begin{tabular}{@{}p{0.30\textwidth}p{0.30\textwidth}p{0.32\textwidth}@{}}
\hline
\textbf{Construction} & \textbf{What it introduces} & \textbf{Invariance it breaks}\\
\hline
Individuation through
$\Gamma=\{\phi=\phibar\}$ &
the cell as an individual, its limit produced from within &
the problem becomes a \emph{free-boundary} problem\\
Spatial assemblage
$\Op_i \mid \Op_o$ &
two heterogeneous operators glued across $\Gamma$ &
\emph{spatial} invariance of the law\\
Temporal assemblage
$\Op^\tau_i=\sum_k a\tau_k\,\Op_{ik}$ &
operators of successive generations superposed and selected &
\emph{temporal} invariance of the law\\
\hline

\end{tabular}
\caption{From physico-chemical kinetics to the living
(Section~\ref{sec:threesteps}). The three constructions successively added to
an invariant kinetics, and the invariances they break. A physico-chemical
kinetics invariant in space and time becomes a differential system that is
\emph{heterogeneous}, \emph{free-boundary} and \emph{only partially autonomous}
--- the residual non-autonomy being the locus at which novelty, and hence
historicity, can enter.}
\label{tab:scheme}
\end{table}

\subsection*{Further remarks}
Below, we outline a few points for future consideration:

- In the theory presented here, the individuation of a membrane $\Gamma$ in an
underlying physical process, together with the creation of an assemblage of
operators, provides a potential explanation for { \it how the dynamics of living systems have evolved from physical dynamics }. Many different kinetic processes can be introduced as operators of the assemblage.

- After individuation each cell thus becomes a { \it  laboratory of novelties } in which new discoveries are tried out.  At the population level, one observes large-scale differentiation and the generation of heterogeneous multiplicities.

- The presence of an exterior operator $\Op_o$ and by the interface it shares with $\Op_i$ allows us to develop an ecological and relational theory of life.. In fact it considers
not only the interior of the cell but also its exterior environment to which it
is coupled. Changes in the environment over time will be no less important than changes in individual morphologies.

- The minimal configuration of heterogenetic assemblages presented here can be
made more complex, since $\Op_i$ and $\Op_o$ need not be single operators, but
can themselves be assemblages of operators with their own constraints. This makes the theory to be open to different ways of assembling and closing constraints.

\subsection{Open problems}

Several questions remain  unanswered.  From the analytical point of view, existence and stability for the heterogeneous assembly can only be established numerically and a free-boundary analysis is still lacking.  

Geometrically, the anisotropic operators of
\eqref{eq:aniso} already produce shapes but the
systematic question of which interfacial morphologies are reachable by which
assemblages, in particular the transition from division to wavy or
labyrinthine interfaces, remains unanswered. 

More generally, the question remains as to what degree of heterogeneity among operators can lead to assembly solutions. So far, we have only considered a  very weak degree of heterogeneity.

We will address these questions in an upcoming publication.

\end{document}